\RequirePackage{fix-cm}
\documentclass[smallextended]{svjour3}       
\usepackage{natbib}

\smartqed  
\usepackage{graphicx}
\journalname{Empirical Software Engineering}
\usepackage{amsmath,amssymb,amsfonts}
\usepackage{algorithmic}
\usepackage{textcomp}
\usepackage{rotating}
\usepackage{multirow}
\usepackage{booktabs}
\usepackage{tabularx}
\usepackage{mdframed}

\usepackage{graphicx}
\usepackage{url}
\usepackage{makecell}

\usepackage{placeins}
\usepackage{fontawesome5}

\usepackage{array}
\newcolumntype{L}[1]{>{\raggedright\let\newline\\\arraybackslash\hspace{0pt}}m{#1}}
\newcolumntype{C}[1]{>{\centering\let\newline\\\arraybackslash\hspace{0pt}}m{#1}}
\newcolumntype{R}[1]{>{\raggedleft\let\newline\\\arraybackslash\hspace{0pt}}m{#1}}

\begin{document}

\newcommand{\AUC}{\textit{AUC}}
\newcommand{\FMEAS}{\textit{F-measure}}
\newcommand{\GMEAS}{\textit{G-measure}}
\newcommand{\MCC}{\textit{MCC}}
\newcommand{\AUCEC}{\textit{AUCEC}}
\newcommand{\RELB}{$RelB_{20\%}$}
\newcommand{\NECM}{$NECM_{15}$}
\newcommand{\RANKSCORE}{\textit{rankscore}}
\newcommand{\RANKSCORES}{\textit{rankscores}}
\newcommand{\RECALL}{\textit{recall}}
\newcommand{\PRECISION}{\textit{precision}}
\newcommand{\ERROR}{\textit{error}}

\newcommand\setrow[1]{\gdef\rowmac{#1}#1\ignorespaces}
\newcommand\clearrow{\global\let\rowmac\relax}
\clearrow

\title{Problems with SZZ and Features: An empirical study of the state of practice of defect prediction data collection}

\titlerunning{Problems in defect prediction data collection}

\author{Steffen Herbold \and Alexander Trautsch \and Fabian Trautsch \and Benjamin Ledel}

\institute{
Steffen Herbold, Alexander Trautsch, and Fabian Trautsch are equally contributing authors. 
\vspace{5pt}\\
Steffen Herbold\\Institute of Software and Systems Engineering, TU Clausthal, Germany\\
\email{steffen.herbold@tu-clausthal.de}
\vspace{5pt}\\
Alexander Trautsch\\Institute of Computer Science, University of Goettingen, Germany\\
\email{alexander.trautsch@cs.uni-goettingen.de}
\vspace{5pt}\\
Fabian Trautsch\\Institute of Computer Science, University of Goettingen, Germany\\
\email{fabian.trautsch@cs.uni-goettingen.de}
\vspace{5pt}\\
Benjamin Ledel\\Institute of Software and Systems Engineering, TU Clausthal, Germany\\
\email{benjamin.ledel@tu-clausthal.de}
}

\date{Received: date / Accepted: date}

\maketitle

\begin{abstract}
\textit{Context:} The SZZ algorithm is the de facto standard for labeling bug fixing commits and finding inducing changes for defect prediction data. Recent research uncovered potential problems in different parts of the SZZ algorithm. Most defect prediction data sets provide only static code metrics as features, while research indicates that other features are also important.

\textit{Objective:} We provide an empirical analysis of the defect labels created with the SZZ algorithm and the impact of commonly used features on results. 

\textit{Method:} We used a combination of manual validation and adopted or improved heuristics for the collection of defect data. We conducted an empirical study on 398 releases of 38 Apache projects.

\textit{Results:} We found that only half of the bug fixing commits determined by SZZ are actually bug fixing. If a six-month time frame is used in combination with SZZ to determine which bugs affect a release, one file is incorrectly labeled as defective for every file that is correctly labeled as defective. In addition, two defective files are missed. We also explored the impact of the relatively small set of features that are available in most defect prediction data sets, as there are multiple publications that indicate that, e.g., churn related features are important for defect prediction. We found that the difference of using more features is not significant.

\textit{Conclusion:} Problems with inaccurate defect labels are a severe threat to the validity of the state of the art of defect prediction. Small feature sets seem to be a less severe threat.
\end{abstract}


\maketitle

\section{Introduction}
\label{sec:introduction}

Defect prediction is an active direction of software engineering research with hundreds of publications. The systematic literature review by \cite{Hall2012} already found 208 studies on defect prediction published between 2000 and 2010, many more have been published since then. Many of these studies were enabled by the sharing of data, highlighted by the early efforts from the PROMISE repository~\citep{promiserepo}, which is nowadays known as Seacraft~\citep{seacraftrepo}. Only few publications on defect prediction collect new data. Instead, most researchers rely on well-known data sets, e.g., the NASA data~\citep{mdp}, the SOFTLAB data~\citep{Turhan2009}, or the data about Java projects from \cite{Jureczko2010} often referred to as PROMISE. A recent literature review on cross-project defect prediction highlights that these and other data sets have become the de facto standard for defect prediction research~\citep{Hosseini2017}. While sharing and re-using data is a good thing in general, heavy re-use may also lead to problems with the external validity of results~\citep{Trautsch2017}. The biggest problem outlined by \cite{Trautsch2017} is that if there are problems that affect the validity of shared data, these problems lead to threats to the validity of all research that re-used this data. Unfortunately, there is evidence that shared defect prediction data is affected by two problems: 1) problems with the defect labels; and 2) limitations regarding the features used by researchers.

The first problem is related to the defect labels, that were determined by different publications that consider different aspects of the defect labeling process. The focus of these publications is mostly on the SZZ algorithm~\citep{Sliwerski2005}, which was applied by most of the currently used data sets (see Section~\ref{sec:related-work}).\footnote{Please note that we use the term SZZ for both phases of SZZ, i.e., the issue linking and the identification of inducing changes. While we are aware that parts of the research community restrict the term SZZ to the identification of inducing changes, i.e., only the second phase of the original SZZ algorithm, this has not been the case in the literature on defect prediction data. Here, the opposite is the case and many data sets use the term SZZ and only apply the first phase of SZZ (see Section~\ref{sec:related-work}).} Research revealed several problems with SZZ, e.g., due to ignoring the affected version field of issue reports~\citep{daCosta2017} or the identification of irrelevant changes~\citep{Mills2018}. The use of a six-month time frame for the assignment of defects to releases has also recently been identified as a problem~\citep{Yatish2019}. Moreover, SZZ relies on the correct labeling of issues as bug by the developers in the issue tracking system. However, research shows that about 33\% of bug reports are mislabeled and are actually improvements or other issues, like outdated documentation~\citep{Herzig2013}. Additionally, SZZ was designed for version control systems that used a mostly linear development process on a main development branch. Due to the success of Git, this is often not the case anymore and there are many new challenges that need to be considered~\citep{Bird2009}. For example, prior research found that data, which takes branches into account, leads to better results~\citep{Kovalenko2018}. 

The second problem with the re-use of the existing data is the limited feature space that researchers use to create defect prediction models. If a data set does not contain certain features, it is unlikely that they are added by other researchers, even if research indicates that these features may be useful. For example, multiple publications indicate that features based on code changes potentially outperform static metrics as features~\citep{Moser2008, DAmbros2012}. Regardless, researchers mostly rely on data sets that only consist of static features~\citep{Hosseini2017}. Thus, many publications are using a potentially inferior set of features, which could alter their results. 

Thus, we know from the related work about many separate problems with defect prediction data, especially with respect to the way software artifacts are labeled as defective, but also due to a potential lack of relevant features. However, each of the prior publications on this topic focuses on a single problem with defect prediction data. What is missing is a view on the impact of the problems if they are not considered in isolation, but together. Within this article, we close this gap and consider the following research question. 

\begin{description}
\item[\textbf{RQ:}] What is the overall impact of the numerous problems with defect labeling and the usually small set of features on defect prediction data and the results of prediction?
\end{description}

We provide insights into the quality problems that existing data may have, with a focus on the defect labeling. To this aim, we performed an in-depth analysis of the weaknesses of existing defect labeling strategies with a focus on SZZ. The original SZZ is still commonly used for the labeling defect prediction data (see Section~\ref{sec:related-work}) and, hence, the \textit{state of practice}. We also compare and discuss our results with respect to newer SZZ variants that are \textit{state of the art} and discuss how the results change. We analyze all aspects of the defect labeling process, i.e., the links between commits and issues, the impact of mislabeled issues, the identification of affected files and the inducing changes, as well as the assignment of defects to releases. Additionally, we use the large sample of data we generate to assess the impact of the lack of features on the performance of defect prediction results.

The primary contribution of our article are the following findings with respect to the current state of practice for the generation of defect prediction data. 
\begin{itemize}
    \item About one quarter of the links to defects detected by SZZ is wrong, both due to missed links as well as false positive links. 
    \item We confirm the results by \cite{Herzig2013} and found that for every issue that is correctly labeled as a bug, there are 0.74 mislabeled bug issues. 
    \item Due to the combination of wrong links and mislabeled issues, only about half of the commits SZZ identifies are actually bug fixing and SZZ misses about one fifth of all bug fixing commits.
    \item The assignment of defects to releases based on a six months time frame, as well as based on the affected versions field of issue tracking systems is unreliable. With SZZ and a six months time frame, we found that for  every  file,  that  is correctly labeled as defective, there are roughly two files that are incorrectly labeled as defective, and two files that are incorrectly labeled as non-defective. Moreover, the quality of the data in the affected version field is questionable for mining purposes without prior manual validation, due to many missing, incomplete, or wrong values.
    \item The difference of using many features of different types over using only static features of the source code is not statistically significant when we consider the cost saving potential of classifiers for release-level defect prediction. Moreover, we found that mislabels during training do not have a significant impact, but mislabels in test data can significantly alter evaluation results.
\end{itemize}

As a secondary contribution, we provide all data we generated through manual validation and repository mining to study the state of practice of defect prediction data generation. As a result, we provide a new defect prediction data set with 4198 features, including change metrics~\citep{Moser2008, Hassan2009, DAmbros2012}, and different aggregation strategies~\citep{Zhang2017}. The data set contains defect data for 398 releases of 38 projects from the Apache ecosystem. 



The remainder of this paper is structured as follows. We discuss the state of practice for the collection of defect prediction data with respect to existing data sets in Section~\ref{sec:related-work}, followed by an analysis of problems with the established data collection methods reported in the state of the art in Section~\ref{sec:issues}. Afterwards, we discuss our suggested improvements to the state of practice in Section~\ref{sec:improvements}. Section~\ref{sec:empirical-study} presents the results of our empirical study on defect labeling and the impact of feature sets. We discuss our results in Section~\ref{sec:discussion} and address the threats to the validity of our work in Section~\ref{sec:threats}. Finally, we conclude the article in Section~\ref{sec:conclusion}.

\section{Existing Data Sets}
\label{sec:related-work}

This article is about the state of practice regarding the collection of defect prediction data. Therefore, the related work are articles that collected defect prediction data. Articles that only discuss specific aspects of this data collection are instead discussed together with the problems of the currently established ways for the collection of defect prediction data in Section~\ref{sec:issues}. We discuss the prior defect prediction data sets with respect to the following criteria: 
\begin{itemize}
    \item the number of distinct projects, as well as the number  of releases;
    \item the level of abstraction of the data set, e.g., modules, files, or classes; 
    \item the features that are provided; and
    \item the defect labeling strategy and the labels that are provided. 
\end{itemize}

\begin{table}
\footnotesize
\begin{sideways}
\begin{tabular}{L{1.9cm}L{1.8cm}lL{1.6cm}L{1.8cm}L{2.4cm}L{1.3cm}L{1cm}l}
\textbf{Dataset} & \textbf{\#Projects} / \textbf{\#Releases} & \textbf{Type} & \textbf{Language} & \textbf{Granularity} & \textbf{Features} & \textbf{Defect Linking} & \textbf{Label Type} & \textbf{Year} \\
\hline
NASA & 13 / 13 & PROP & C, C++, Java & Module & SIZE, COM & NA & Binary & 2003 \\
ECLIPSE & 1 / 3 & OSS & Java & File, Package & SIZE, COM, CHURN & SZZ & Counts & 2007 \\
SOFTLAB & 5 / 5 & PROP & C, C++ & Module & SIZE, COM & NA & Binary & 2009\\
PROMISE & 37 / 92 & Mixed & Java & Class & SIZE, COM, OO & REGEX & Counts & 2010 \\
RELINK & 3 / 3 & OSS & Java & Class & SIZE, COM & GOLDEN & Binary &  2011 \\
AEEEM & 5 / 5 & OSS & Java & Class & SIZE, COM, CHURN & SZZ & Counts & 2012 \\
NETGENE & 4 / 4 & OSS & Java & File & SIZE, COM, GEN & SZZ & Counts & 2013 \\
MJ12A & 18 / 70 & Mixed & Java & Class & SIZE, COM, OO, CHURN & NA & NA & 2015 \\
SHIPPEY & 23 / 69 & OSS & Java & Class, Method & SIZE, COM & SZZ & Counts & 2016 \\
GITHUB & 15 / 15 & OSS & Java & Class, File & SIZE, COM, OO, DOC, CLONE & SZZ & Binary & 2016 \\
UNIFIED & 37 / 71 & OSS & Java & Class, File & SIZE, COM, OO, DOC & NA & NA & 2018 \\
RNALYTICA & 9 / 32 & OSS & Java & File & SIZE, COM, OO, CHURN & Affected Version & Counts & 2019 \\
\hline
JIT & 6 / - & OSS & Java, Perl, C++, Ruby & Commit & CHURN, DEV & SZZ & Binary & 2013 \\
AUDI & 3 / - & Prop. & C & Commit & SIZE, COM, CHURN & SZZ & Binary &  2015 \\
MT & 2 / - & OSS & Python, C++ & Commit & CHURN, REVIEW & SZZ & Binary & 2017 \\
FJIT & 10 / - & OSS & Java, JavaScript, Perl, C & Commit+File & CHURN, DEV & SZZ & Binary & 2019 \\
\hline
BUGHUNTER & 15 / - & OSS & Java & File, Class, Method & SIZE, COM, OO, CLONE & SZZ & Binary & 2020 \\
\hline
\end{tabular}
\end{sideways}
\vspace{1pt}
\caption{Overview of existing public data sets for defect prediction research. The data sets in the first compartment are for release-level defect prediction, the data sets in the second compartment are for just-in-time defect prediction. The BUGHUNTER data set contains release-level data for all bug fixing commits.}
\label{tbl:datasets}
\end{table}


Overall, we are aware of fifteen publicly available data sets for defect prediction as of August 2019. The NASA~\citep{mdp}, ECLIPSE~\citep{Zimmermann2007}, SOFTLAB~\citep{Turhan2009}, PROMISE~\citep{Jureczko2010}, RELINK~\citep{Wu2011}, AEEEM~\citep{DAmbros2012}, NETGENE~\cite{Herzig2013a}, MJ12A~\citep{Madeyski2015}, SHIPPEY~\citep{Shippey2016}, GITHUB~\citep{Toth2016}, UNIFIED~\citep{Ferenc2018, Ferenc2020a}, RNALYTICA~\citep{Yatish2019} contain release-level data and the JIT~\citep{Kamei2013}, AUDI~\citep{Altinger2015}, MT~\citep{McIntosh2018}, and FJIT~\citep{Pascarella2019} just-in-time data. The BUGHUNTER~\citep{Ferenc2020} contains release-level data for all bugfixing commits, i.e., is neither a traditional release level data set, nor a just-in-time data set.
Table~\ref{tbl:datasets} gives an overview about the data sets. The data contains mostly open source projects (OSS), but also proprietary projects (PROP). The PROMISE data is actually a mix of 15 open source projects with 48 releases, 6 proprietary projects with 27 releases, and 17 student projects with 17 releases. 

These data sets can be distinguished between data for release-level defect prediction and just-in-time defect prediction. There are two major differences between the release-level and just-in-time data sets. 1) Release-level data contains features for all software artifacts for a certain revision (usually a release) of a project, while just-in-time data contains data for every commit of a project, possibly restricted to the main development branch. 2) The release-level data consists mostly of features that measure the source code directly, e.g., its size, structure, or coupling. Just-in-time data consists mostly of features that measure the changes, e.g., the number of lines that are changed. The second difference has a major consequence regarding the programming languages that are considered: the collection of data about the source code structure requires language-specific tooling for the (static) analysis of source code, while the collection of data about code changes and ownership can be done directly using the version control system. This is reflected directly in the languages of the projects: while most release-level data sets are only for one specific programming language, two out of three just-in-time data sets are for a diverse set of languages. 

We note that there is a strong focus on Java in the release-level data sets. Although we have no scientific evidence for this, we believe that the reason for this is likely the good tool support for the static analysis of Java. Moreover, we note that the features for the release level data sets are mostly static product metrics of the types that measure the size (SIZE), code complexity (COM), or aspects of object orientation (OO), e.g., using Chidamber and Kemerer's metrics~\citep{Chidamber1994}. The GITHUB, UNIFIED, and BUGHUNTER data sets also contain other features based on static product metrics, i.e., regarding the code documentation (DOC) and code clones (CLONE). The notable exceptions are the AEEEM, MJ12A, and RNALYTICA data, which also contain features based on code changes, e.g., added and deleted lines (CHURN). The AEEEM data even contains features that measure the entropy of code changes as proposed by \cite{Hassan2009} and \cite{DAmbros2012}. The ECLIPSE, AEEEM, and MJ12A data also contain metrics regarding prior bug fixes. The MJ12A data is an extension of a subset of the PROMISE data with CHURN metrics. The UNIFIED data is a special case: this data set is actually a combination of the PROMISE (only OSS projects), AEEEM, ECLIPSE, and GITHUB data. All data is conserved as is in the data set and augmented with additional metric data to create the UNIFIED data set~\citep{Ferenc2018, Ferenc2020}. Most data sets for release-level defect prediction are using the file, class, respectively module level. These are all very similar, as they encompass the complete contents of a single file in most cases, with the exception of anonymous and inner classes, that can lead to differences. The notable exceptions are the SHIPPEY and BUGHUNTER data, which also contain method-level data.

For the just-in-time defect prediction data, we note that JIT, MT, and FJIT are independent of the programming language, which is a big advantage for generalizing the defect prediction approach. These data sets are using metrics that can directly be inferred from the version control system. MT also uses metrics about code REVIEW. An important difference between the data sets is that the JIT and MT data only contain the information which commits induced a defect, while  FJIT contains information which changes to files (hereafter referred to as file actions) induced a defect. The AUDI data set is not comparable to the other two data sets: the data is about source code that was not written by developers, but instead generated from Simulink models developed by engineers. Moreover, the data contains not only information that is collected from the version control system, but also static product metrics about SIZE and COM. 

The identification of defective artifacts is a major aspect of defect prediction data that can greatly affect the quality of the data. For example, \cite{Yatish2019} recently found that labeling based on the affected releases leads to large differences in comparison to labeling based on keywords and a six-month time window as proposed by \cite{Fischer2003}. The de facto standard approach in research is the SZZ algorithm~\citep{Sliwerski2005} as it was used by \cite{Zimmermann2007}. The SZZ algorithm first identifies bug fixing commits and then the corresponding inducing changes. However, the identification of inducing changes is, to the best of our knowledge, so far ignored by all release-level data sets. Instead, only bug fixing commits are identified using SZZ and then the six-month time window as proposed by \cite{Fischer2003} is used by \cite{Zimmermann2007}. This approach was used for the collections of the ECLIPSE, AEEEM, NETGENE, SHIPPEY and GITHUB data sets. The BUGHUNTER data uses the links on GitHub between commits and issues to determine bug fixing commits. To determine which methods, classes, and files contribute to the bug fix, they use a SZZ variant similar to the work by \cite{Williams2008} that determines ranges of lines based on diffs and match these ranges to the code artifacts. The authors manually validated this approach and find that this works well in over 99\% of the cases. We note that the manual validation only covered if changed source code was matched correctly, and not if the changes contributed to the bug fix. 

The just-in-time defect prediction data sets JIT, AUDI, MT, and FJIT data\footnote{According to the replication kit, the FJIT data may use only a subset of SZZ for identification of bug fixing commits, i.e., a pure keyword based approach.} use SZZ including the identification of inducing commits, usually with the addition to ignore white-space and comment only changes. The RELINK data was created as a case study for an issue linking approach. The authors created manually validated issue links and used these links for the identification of bug fixing commits. All files that were implicated in any bug fix are considered as defective, without using a time window. The PROMISE and MJ12A data identify bug fixing commits using regular expressions applied to commit messages and a six-month time window. For the NASA and SOFTLAB data, no information on how the defects are linked to source code is given. Since the UNIFIED data is an aggregation of other data sets that reuses defect labels, there is no defect identification approach. The RNALYTICA data uses an approach for the identification of defects based on the affected version field of the bug tracking system. The authors establish links between commits and issues based on references from issues to the commits in which they were addressed. They then use the hunks changed in these commits to identify which files were changed. The authors then use the affected version field of the issue tracker to assign the defect for the file to releases. 

We note that there may be additional data sets, that we did not discuss above. For example, we excluded the data used by \cite{Zhang2014, Zhang2017} based on the census data by \cite{Mockus2009}. The reason for this exclusion is that this data is, to the best of our knowledge, not publicly available anymore, because the links in both papers do not work anymore. Regardless, these data sets would not add anything regarding the methodology for collecting defect prediction data, as the approach is almost exactly the same as for PROMISE and MJ12A: the defect identification is based on commit messages and a six month time window, the metrics are SIZE and COM, and in case of~\citep{Zhang2014} also CHURN. Data like Defects4J is out of scope of our discussion of defect prediction data. While Defects4J contains detailed information about a subset of defects from a project, the data neither contains features, nor examples for changes without defects. As a consequence, such data would only consist of positive labels and not be suited for any defect prediction experiment. 

\section{Problems with Existing Data Sets}
\label{sec:issues}

The sharing of public data sets in defect prediction is a success story that enabled defect prediction researchers to conduct many experiments with the data. Moreover, the use of the same data by different authors enables comparisons between approaches through meta studies, as was, e.g., done by \cite{Hall2012} who exploited that many papers are based on the NASA data. However, there are a number of potential problems that researchers found regarding the data collection procedures used for the creation of the defect prediction data sets. Within this section, we summarize problems regarding algorithms for defect labeling and the features available in current data sets. 

\subsection{Defect Labeling}
\label{sec:label-issues}

Defect labels are the key component of any defect prediction data set. These labels mark artifacts as defective, e.g., files in a release or in a commit. These labels are the dependent variable that defect prediction models try to predict based on the independent variables, i.e., the features. In the existing defect prediction data, labels are either binary or defect counts. Noisy defect labels may negatively affect the training of defect prediction models or make the evaluation of results unreliable. Especially the impact of the noise on the evaluation of results is problematic. For example, if a defect labeling approach marks too many instances as defective, i.e., produces false positives, the commonly used measures \RECALL, \PRECISION{} and \FMEAS{} are not trustworthy anymore, because values may change if the distribution between defective and non-defective instances changes. Consider an example with 100 software artifacts, 25 artifacts are actually defective, but the defect labeling algorithm introduces noise and labels 50 artifacts as defective. A trivial model that predicts everything as defective will overestimate the \PRECISION{} as 0.5 instead of the actual 0.25, which would also affect the \FMEAS{} which would be 0.7 instead of 0.4. 

As we discussed above, the de facto standard for labeling defective instances is the use of the SZZ algorithm~\citep{Sliwerski2005} in the variant used by \cite{Zimmermann2007}. The SZZ algorithm works in two steps. First, bug fixing commits are identified. The SZZ algorithm tries to find a matching issue, based on the numbers found in the commit messages. In case any number is found, the algorithm tries to find an issue for the project that has the same number. If an issue is found, semantic checks for the following properties are performed~\citep{Sliwerski2005}:
\begin{itemize}
    \item The issue was resolved as FIXED at least once.
    \item The author of the commit is assigned to the issue.
    \item The title or description of the issue is contained in the commit message.
    \item One or more files that are changed by the commit are attached to the issue. 
\end{itemize}

A commit is identified as bug fixing if there is at least one linked issue, that passes at least two of the above semantic checks. In case only one semantic check is passed, the commit is labeled as bug fixing, if the commit message contains a term like ``bug'' or ``fix'', or it is clear that a number in the commit is a link to an issue, e.g., because the number starts with ``Bug \#111'', or the commit contains only a list of numbers. 

Once a commit is identified as bug fixing, the second part of the SZZ algorithm is the identification of the inducing changes. SZZ identifies the last changes to each line that was touched as part of a bug fixing commit as candidate for an inducing change. All candidates, that took place before the reporting date of the issue are immediately considered as bug inducing changes. Changes that took place after the reporting date of the issue are suspect, because they were performed after the bug was already in the software. However, because of the chance of bad fixes or partial bug fixes, the changes are not automatically discarded. If the suspects are part of a bug fixing commit (partial fix) or the commit contains changes that are inducing for a different bug (weak suspect), they are considered as inducing. The remaining suspects are considered to be hard suspects and not inducing for the bug fix. The identification of the bug inducing commits is only used for the just-in-time data. For the release level data sets, defects are assigned to releases based on the reporting date: all bugs that were reported in the first six months after the release are assigned to a release~\citep{Zimmermann2007}. 

While our focus is on defect prediction data, a study by \cite{RodriguezPerez2018} further highlights the relevance of SZZ beyond defect prediction data. The study highlights that the original SZZ is commonly used, i.e., 38\% of the identified literature. Another 40\% of the identified literature refers to some adoption of SZZ, without specifying what this adoption is. Only 14\% of the publications that use SZZ specifically mentioned that they use a variant that ignores cosmetic changes, like documentation changes.

In recent years quality problems with the labels produced by SZZ came into focus. The impact of using all changes in a bug fixing commit as foundation for the defect labeling was investigated in detail by \cite{Mills2018}. The authors manually validated which files that were modified in a bug fixing commit were actually part of the bug fix and found that about 64\% of file changes made in bug fixing commits are not part of the bug fixes, but other changes. They found that mistakenly identified files are due to code that is only added and not modified or deleted (46.58\% of all cases), changes are performed on test code (30.90\%), refactorings (8.73\%), and changes of comments (8.49\%). SZZ already ignores pure additions for the identification of inducing changes, because there is no prior commit, where the code was last changed. To the best of our knowledge, none of the SZZ implementations used to create the defect prediction data sets ignores test code, refactorings, or comments. This means that based on the estimation of \cite{Mills2018}, about 34\% of files in bug fixing commits are false positives, i.e., incorrectly identified as defective. 

Another potential source of false positives of SZZ are commits that are mistakenly identified as bug fixes. With SZZ, there are two main sources for this problem: the first is due to the strategy for the identification of links between commits and issues, that works based on numbers. If a core developer of a project fixes a bug with an issue number like one, 256 or other frequently occurring numbers, every commit by this developer that contains this number will be identified as a bug fixing commit. While \cite{Bird2009a} found that this problem can be mitigated through additional filters, e.g., based on the date of the commit and the issue resolution, this is not part of the standard SZZ algorithm. There are also approaches that try to recover links between commits and issues, that are not explicit, e.g., ReLink~\citep{Wu2011}. Such links also cannot be captured by SZZ. The second source of false positive for bug fixing commits are the issues themselves. According to \cite{Herzig2013}, about 33\% of issues that are reported as a bug in the issue tracking system are actually requests for new features, bad documentation, or simply result in refactorings. They found that due to this 39\% of the identified defective files were actually not defective. A smaller study on this topic was conducted by \cite{Antoniol2008}, who came on a smaller sample to a similar result.\footnote{Unfortunately, \cite{Antoniol2008} did not specify their criteria for the differentiation between bugs and enhancements and only stated that bugs are corrective changes. Due to the uncertainty of this term, we only compare our results to \cite{Herzig2013} in the following, as they provide clear guidelines for the identification of different issue types.} To the best of our knowledge, only the NETGENE data is based on manually validated issues. That both happen in practice can, e.g., be seen with the issue NUTCH-1.\footnote{https://issues.apache.org/jira/browse/NUTCH-1} This test issue created by a core developer was not for any real bug in the software. However, all commits by this developer for the Apache Nutch project, where the message contains the number one will be mistakenly identified as bug fixing. Moreover, \cite{Herzig2013} note that these mislabels are mostly irrelevant for developers, but very important for data miners: 
\begin{quote}
``\textit{The question of whether an issue is a bug or not is a hard one. And the definition of a bug not only differs between users and developers but also between developers themselves. In principle, if an issue bothers the user, the developer should fix it, whether or not he considers it to be a bug or not. From this perspective, the issue report category does not matter. But a data miner building a defect prediction model must distinguish between bugs and non-bugs. Otherwise, the prediction model would predict changes, not defects.}''~\cite{Herzig2013}
\end{quote}

Another aspect related to this is to differentiate between \textit{intrinsic} and \textit{extrinsic} bugs, a notion recently introduced by \cite{Perez2020}. This notion extends the concept that not all bugs have inducing changes \citep{RodriguezPerez2018a}. Intrinsic bugs are ``introduced by one or more specific changes to the source code''~\citep{Perez2020} and extrinsic bugs were introduced, e.g., ``from external requirements or changes to the requirements''~\citep{Perez2020}. In the sense of \cite{Herzig2013} and our study, we only consider intrinsic bugs, because issues due external dependencies and changing requirements are not bugs, but feature requests for improvements. Thus, extrinsic bugs are not within the scope of defect prediction. Regardless, it is valid for the developers to label extrinsic bugs as bugs within the issue tracker. Thus, both the work by \cite{Herzig2013} and \cite{Perez2020} indicate that there are likely many issues mislabeled as bug in issue trackers, at least from the point of view of defect prediction that is interested in intrinsic bugs. 

The assignment of the identification of the inducing file actions has also come under scrutiny. \cite{daCosta2017} investigated how well SZZ identifies inducing commits and found that SZZ implementations perform better, if they ignore changes that only modify whitespaces. This is in line with the findings by \cite{Mills2018}. Moreover, they suggest that using the affected versions field of issue tracking systems can improve the validity of SZZ results. Developers can use this field to mark versions of a software that are affected by a specific defect. Regardless, all data sets we discussed in Section~\ref{sec:related-work} use a basic SZZ variant that does not ignore whitespace changes or use the affected version field. However, \cite{daCosta2017} do not suggest how the affected version should be integrated into the SZZ algorithm. \cite{Perez2020} went one step further with their investigation of the bug inducing changes of SZZ and manually validated the inducing changes of 86 bugs for two projects.\footnote{These are the intrinsic bugs from their article.} They found that about 70\% of the correct inducing changes can be found.\footnote{76\% for the Nova project, 63\% for the ElasticSearch project} However, they also find that of all inducing changes identified by SZZ only 24\% are actually true positives. However, they only consider the first change that is related to the introduction of the bug as inducing. Additional changes that are made later, which may have also contributed to the bug are also identified as false positive. Regardless, these findings indicate that SZZ may overestimate the number of bug inducing commits. 

\cite{Yatish2019} directly used the affected version field to assign defects to releases. Theoretically, this could lead to a perfect assignment of defects to releases. However, this depends on the maintenance of this field in the issue tracker by the developer. In practice, the value of this field is usually set by the reporter of an issue as the version of the software that the reporter currently uses. An analysis if this defect was already in the software in earlier versions is often not performed and, consequently, the field is not updated. An example for this is the issue CAY-1657.\footnote{https://issues.apache.org/jira/browse/CAY-1657} The affected version of that issue is 3.1M3. However, as part of the description, the author writes ``\emph{I am sure this affects ALL versions of Cayenne, but my testing is done on 3.1 M3/M4}''. That this is not an isolated problem is also suggested by further anecdotal evidence, e.g., on the Apache Spark mailing list.\footnote{https://hdl.handle.net/21.11101/0000-0007-E183-6} Additionally, the affected version field is often not used at all. For example, \cite{daCosta2017} report that only 1,268 of the 32,033 linked bug reports contained data in the affected versions field, i.e., less than 4\% of the issues. Thus, this approach is likely to lead to many false negatives, i.e., mistakenly not assigned defects to releases that are affected, because the field is not maintained properly.

Regarding the six month time frame that was proposed by \cite{Zimmermann2007} for the release assignment with SZZ, we could not find an empirical basis for this in the literature. \cite{Yatish2019} already broke with this rule and demonstrated that there are both defects within the six-month time that were introduced after the release, leading to false positives, as well as defects that were fixed more than six months after the release, leading to false negatives. Thus, the work by \cite{Yatish2019} provides a strong indication that the complete history of a project after the releases should be considered for assigning defects to releases. We are not aware of other studies that evaluate the impact of the six month time frame. 

Finally, the mislabels in data we discuss above may be acceptable, if they do not affect the prediction models. The literature indicates that the there is a threat to the validity of the results due to this kind of noise in the data. According to \cite{Herzig2013}, the mislabeled issues translate into bad estimation of defect proneness of files, because 16\% to 40\% of the top 10\% of files with most defects change due to the correction of issue types. Additionally, \cite{Tantithamthavorn2015} found that while the models trained on noisy data may achieve similar \PRECISION{} as models trained without noise, their \RECALL{} is typically reduced by 32\%-44\%, suggesting a degradation in prediction performance due to the noise. We note that the studies by \cite{Herzig2013} and \cite{Tantithamthavorn2015} both only consider mislabels due to wrong issue types. The other potential reasons for mislabels are not considered. 

\subsection{Incomplete Feature Sets}
\label{sec:lack-of-features}

The features (or independent variables) are at the core of every learning algorithm: they are the information that is available to make a decision or they can be correlated with the outcome. If good features are missing in defect prediction data, this can lead to a loss in performance. This loss in performance can vary between algorithms used for training prediction models. This leads to a troubling question: if researchers find a difference between defect prediction approaches on data that does not contain all important features, would the difference still be there if all relevant features are available? As a consequence, conclusions regarding the performance of defect prediction algorithms based on data that does not contain features that were demonstrated to be valuable have a severe problem with the external validity of the findings. We note that the curators of data sets are not to fault for features that are not within their data sets. Most data sets are collected based on what is currently known in the state of the art and contain the currently relevant features. However, this problem can easily manifest if the data is used for years and, in the meantime, the state of the art progresses and suggests that important features may be missing from older data sets. The defect prediction literature suggest that there are at least several such kinds of such features: CHURN related features, as well as different variants of aggregated features. 

CHURN related features are based on the findings that defects are more likely in often changed parts of a software, especially in case a prior change already removed defects~\citep{Rahman2011}. Moreover, such features may also include information about past defects (BUG), e.g., the number of prior defects that were already corrected in a file. Publications that include CHURN features consistently find that CHURN features are among the most important predictors for defects. This already started with the pioneering work by \cite{Ostrand2005} at Bell labs and was later confirmed, e.g., by \cite{Moser2008} and \cite{DAmbros2012}. \cite{Hassan2009} proposed to use the concepts of entropy and linear decay to further improve the impact of CHURN features, which was also confirmed by \cite{DAmbros2012}. Another strong indicator for the importance of CHURN related features is that they are the main features in the just-in-time defect prediction data sets in combination with code ownership. Regardless, the only release-level data sets that contain CHURN features are ECLIPSE, AEEEM, MJ12A, and RNALYTICA. 

However, there is another problem that is known related to CHURN features, i.e., those that take the history into account. The current data sets collect this data only based on the main development branch of a repository. However, due to the advent of Git as version control system, features are often developed on branches. These feature branches are merged through a single merge commit into the main development branch. If only the main branch is considered, information about the history of the development is ignored. Only the RNALYTICA data may consider
branches for the CHURN feature, as the authors state that ``one must focus on the development activities of interest that correspond to the release branch''~\citep{Yatish2019}. However, \cite{Yatish2019} do not discuss further how branches are handled, which is why we are not sure if and how different branches affect the collection of the CHURN metrics or if the branches are used for the assignment of defects to releases. To the best of our knowledge, the other current defect prediction data sets only use the main development branch. \cite{Kovalenko2018} evaluated the impact of using feature branches as part of the data collection on results of various software mining approaches, including defect prediction. They found that the performance of defect prediction may improve slightly, if data from branches is included in the mining process. In particular, they found that the results are never worse. Thus, while this problem probably does not have the same impact as the general lack of CHURN features, this could still lead to underestimating the performance of defect prediction approaches.

There are other types of features, which are ignored by current defect prediction data altogether. \cite{Zhang2017} found that how measurements from lower level software artifacts are aggregated into metrics for higher level artifacts impacts the performance of defect prediction models. A popular example of such a metric is the Weighted Method per Class (WMC) metric from the Chidamber and Kemerer's metrics for object-oriented software~\citep{Chidamber1994}. This metric measures the complexity of a class by summation of the complexity of the methods. However, \cite{Zhang2017} found that aggregation through a single marker, like summation, can actually lead to inferior defect prediction models. Instead, they recommend to use different aggregation strategies to provide multiple aggregations such as summation, median, and the standard deviation and later use feature reduction techniques to remove redundancies. We note that while \cite{Zhang2017} found that using all aggregation schemes leads to the best results overall, they also observed that using only summation is a close second, i.e., the advantage of using multiple summation schemes may be negligible. Regardless, to the best of our knowledge, none of the current publicly available data sets support this kind of analysis. 

\cite{Plosch2008} analyzed the correlation between warnings produced by the static analysis tools FindBugs\footnote{http://findbugs.sourceforge.net/} and PMD.\footnote{https://pmd.github.io/} They found that the warnings have a stronger correlation with bugs than OO and SIZE metrics. In contrast, \cite{Rahman2014} found that features from static analysis tools do not improve the performance of defect prediction models. Due to the contradictory results, we believe that more data is required, e.g., through additional studies that evaluate the impact of such features. \cite{Bird2011} found that code ownership is correlated with defects and may be used to improve defect prediction models. \cite{Palomba2019} found that it may be useful to derive code smells from static metrics, to improve defect prediction models. We note that this data need not be actually part of the data sets, but can be derived as long as static source code metrics like number of methods are available. \cite{Bowes2016} found that dynamically collected features from mutation testing can also improve defect prediction models. A notable difference between the mutation features by \cite{Bowes2016} and all other features we discuss is that they require that the software can be compiled and executed. Moreover, \cite{Thongtanunam2016} found a correlation between code review activity and defects, however, they did not study if this correlation can improve defect prediction models. Similarly, \cite{Spadini2018} found that smelly test cases are correlated with defects, but have also not studied if this correlation can improve prediction models.

\section{Improving Defect Prediction Data Collection}
\label{sec:impr-collection}

We now outline how we believe that defect prediction data can be improved to avoid the problems discussed in the literature. Table~\ref{tbl:summary-of-problems} summarizes the key problems we discussed in Section~\ref{sec:issues} as well as the solutions we suggest within this section. In the following, we describe the improvements in detail. Wherever possible, we re-use or adapt existing solutions from the literature and rather propose to combine the separate findings and solutions.

\begin{table}[]
\centering
\begin{tabular}{p{3cm}p{5.5cm} l}
\textbf{Problem} & \textbf{Improvement} & \textbf{Identifier} \\
\hline \hline
Wrong links to bugs & Exploit Jira identifier pattern & JL \\
 & Manually validate correctness of links & JLM \\
Wrong issue types & Manually validate issue types & JLMIV \\
Too many inducing changes & Exclude non-production code, whitespace and documentation changes & JLMIV+\\
& Ignore refactorings & JLMIV+R\\
6-month time-window & Use affected versions field  & JLMIV+RAV\\
& Use commit graph & IND-JLMIV+R \\
Lack of features & Provide a large set of features & - \\
\hline
\end{tabular}
\caption{Summary of problems with defect prediction data creation and the improvements we consider. The problems are described in detail in Section~\ref{sec:issues}, the improvements are described in Section~\ref{sec:impr-collection}. The identifiers refer to the SZZ variant that implements the improvement.}
\label{tbl:summary-of-problems}
\end{table}

\subsection{Improving Defect Labeling}
\label{sec:improvements}

In principle, we believe that the SZZ algorithm~\citep{Sliwerski2005} provides a very good foundation for the labeling of defects. Thus, we do not propose a radically new algorithm, but rather modify the SZZ algorithm to work well together with the Jira issue tracking system, as well as take the problems from the state of the art into account. 

As we described in Section~\ref{sec:label-issues}, the SZZ algorithm may suffer from misidentified defect links due to often occurring numbers like 1, that are not related to an issue number. We note that SZZ was designed with the Bugzilla issue tracker in mind. Here, issues identifiers are just a single number, i.e., there is no good resolution for this. This is different in Jira, where the issue identifiers have the structure $<$PROJECTID$>$-$<$NUMBER$>$. Thus, we modified the identification of linked issues to take this structure into account to define a new linking approach we call JL\footnote{Short for Jira Links}. JL is conceptually identical to SZZ, the difference is the focus on Jira instead of Bugzilla.

JL exploits the semantics of the string descriptor of Jira, i.e., we search for the complete identifier in commit messages, and not just any number. The drawback of this is that spelling problems in the project identifier would mean that we miss issue links. To account for this, we manually check all strings that are a combination of a string followed by an integer and supply a list with all wrong spellings, such that they can be corrected by the linking algorithm. While this requires manual effort, this can be done in a matter of minutes. The problem with JL is that it also captures links to commits, where an issue is only mentioned, but not actually addressed. Moreover, if the numbers are alone, i.e., not part of a Jira identifier, JL may miss links. SZZ can detect such links, because the algorithm works only on numbers. Thus, to account for links that SZZ detects but JL misses, we semi-automatically analyze all messages of commits that contain a link determined by JL or SZZ. The goal of this additional step is to create a validated set of links from commits to issues. For many commits, this is not a problem. In case we determined only one link from a commit to an issue, and this link is established because the exact name of the issue occurs at the start of the commit message, we assume that this commit addresses the mentioned issue. An example for such a commit message from the ant-ivy project is ``\emph{IVY-1391 - IvyPublish fails when using extend tags with no explicit location attribute}''. An expert must inspect all remaining commit messages for which a link was detected regarding two criteria: 1) are the links correct, i.e., are the issues actually mentioned by the commit message and 2) which issues were actually addressed by the commit and which issues were only mentioned. Only correct links that were actually addressed in the commit are then validated by the expert. We refer to the combination of JL with the validated data as JLM\footnote{Short for Jira Links Manual}. 

For both JL and JLM we use rules similar to SZZ~\citep{Sliwerski2005} to determine if a commit is bug fixing: 
\begin{itemize}
    \item a bug fixing commit must have a validated link to an issue that is validated as BUG; and
    \item the linked issues must have been in a closed or resolved state at any point in its lifetime. 
\end{itemize}

The major assumption behind JLM is that the labeling of issues as bugs in the issue tracker is correct. Since we know from the results by \cite{Herzig2013} that this is often not the case, we propose that the type of issues should be manually validated. Taking pattern from \cite{Herzig2013}, we used the following five categories. 
\begin{itemize}
    \item BUG for null pointer exceptions, runtime or memory issues caused by defects, or semantic changes to the code to perform corrective maintenance task. This is the same as the BUG category from \cite{Herzig2013}.
    \item IMPROVEMENT for feature requests or the non-corrective improvement of existing features. This bundles the categories RFE (Feature Request), IMPR (Improvement Request), REFAC (Refactoring Request) from \cite{Herzig2013}.
    \item TEST for issues that only require changes to the software tests. This category was not used by \cite{Herzig2013}.
    \item DOC for requested changes to the documentation of the software. This is the same as the DOC category from \cite{Herzig2013}.
    \item OTHER for all other issues, e.g., questions or brainstormings. This is the same as the OTHER category from \cite{Herzig2013}.
\end{itemize}

Our reasons for the differences between our work and \cite{Herzig2013} are mainly due to the efficiency, i.e., a faster manual validation process. The different types of IMPROVEMENT are often very hard to distinguish based on the description of an issue, while they all lead to improvements of the software. We kept the DOC and OTHER and added TEST because these are clearly distinguishable from the other issue types. For maximal efficiency, one could also use a simple binary classification, i.e., BUG, and not BUG. For our research, we decided against this to facilitate research using this data regarding the automated correction of issue types. 

We propose that the manual validation should be done in two steps. First, all linked issues of type BUG should be independently labeled by two experts. The experts have access to the description and comments of the issue, as well as the source code that was changed as part of the commits that were linked to this issue. If both experts agree, we assume their assessment is correct. In case of disagreement, the issues should be presented to a panel of at least two experts, one of which did not participate in the initial labeling. The experts then decide the issue type based on the blinded labels determined by the two experts, the issue description and comments, and the source code changes. This validation should be based on the principle ``innocent unless proven guilty'', i.e., in case there is doubt whether the issue is a BUG or not, the experts should not modify the label, i.e., always label such issues as BUG. 

While other issues of type other than BUG could also be manually analyzed, the work by \cite{Herzig2013} showed that bugs are almost always correctly labeled as BUG. Thus, we suggest to restrict the manual labeling to issues of type BUG, due to the time intensive nature of the manual labeling step. In the following, we use JLMIV\footnote{Jira Links Manual Issues Validated} to refer to bug fix labeling that accounts for the manual validation of issue types. 

The results by \cite{daCosta2017} and \cite{Mills2018} indicate problems with the way SZZ determines bug inducing changes. Both studies highlight that changes that only affect the whitespaces or modify comments should be ignored during the identification of bug inducing changes. To address this concern, we use a regular expression approach to identify changes that only modify comments or whitespaces and ignore them, similar to what \cite{Kim2006} proposed. \cite{Mills2018} also found that changes to tests are also inadvertently covered during the search for bug inducing changes. Since bugs can, by definition, only be in production code, these changes are all false positives. We extend this notion to changes to examples or tutorials, which may contain code files. These are also not production code, but documentation of the project. We are not aware of an SZZ variant that takes this into account. Additionally, we ignore changes that can be explained by refactorings similar to the work by \cite{Neto2018, Neto2019}. Based on the results by \cite{Mills2018}, these modifications should be able to account for about 94.6\% of the false positive bug inducing changes. We refer to this improvement of the detection of inducing changes as JLMIV+R. 

\cite{daCosta2017} also noted that SZZ should take the affected version field of issue tracking systems into account, because this gives further information about the time when the software was defective. In their study, \cite{daCosta2017} mark all changes after the release of the earliest affected version as incorrect. \cite{Yatish2019} assign bug fixes directly to releases based on the affected version field and do not use SZZ at all. In comparison to \cite{Yatish2019} and \cite{daCosta2017}, we do not consider the affected version label as ground truth, due to the reasons we discussed in Section~\ref{sec:label-issues}. Because we assume that the affected version field is likely missing completely or at least missing older releases that were also affected by a bug, we believe that the approaches by \cite{daCosta2017} and \cite{Yatish2019} are too strict for real world data and would lead to false negatives, i.e., not assigning bugs to affected releases because the affected versions field is incomplete. A less strict variant that takes the affected version field into account would be to integrate the affected version in the strategy to determine inducing changes of the SZZ algorithm, based on how the bug reporting date in the issue tracker is used. SZZ assumes that changes after the reporting date of a bug are suspect, but may still be inducing for the bug fix, e.g., because they are bad fixes or partial fixes. The same logic can be applied to changes that happen after the release of an affected version. Thus, our proposal to utilize the affected version field to enhance the SZZ algorithm is to extend the notion of suspect changes and use the minimal date of the release of all affected versions and the reporting date of the bug as the boundary for suspects. We refer to this approach as JLMIV+RAV. 

For release level-data, there is another problem to consider, i.e., how we decide which releases were affected by which bugs and label files within releases accordingly. We already discussed that the six month timeframe has no empirical foundation and leads to mislabels, as demonstrated by \cite{Yatish2019}. However, since we believe that the affected version field is unreliable (see sections \ref{sec:label-issues} and \ref{sec:groundtruth-inducing}) , we propose a different approach than \cite{Yatish2019}. We propose an approach that is directly based on the bug inducing changes. If the bug inducing changes are determined correctly, this means that the bug was in the software, when the last non-suspect bug inducing change took place. Suspect changes have to be excluded here, because there is confirmation in the issue tracking system that the bug already affected the system, when this change took place. Similarly, we can determine when the bug was fixed as the last bug fixing commit for the bug. Consequently, a bug affects a release, if all non-suspect inducing changes took place before the release, i.e., there is a path in the commit graph from all inducing changes to the release, and at least one bug fixing commit took place after the release. We refer to this approach as IND-JLMIV. This approach is in line with the theory of how bugs are introduced by \cite{Perez2020}, which is based on the idea that an intrinsic bug is within the software once the changes that lead to the bug are performed. Assuming that all inducing changes are required to produce the bug, the last inducing change is the point in time where the bug was introduced into the software. 

\subsection{Reducing the Lack of Features}

The remedy for the lack of features is straight forward: collect data for more features based on prior work from the state of the art. This requires no change in the general approach for data collection, i.e., to to compute features from repositories mostly using off-the-shelf static analysis tools. The only required change is that more tools are used to collect these features to obtain a broader feature set.

\section{Empirical Study}
\label{sec:empirical-study}

\begin{figure}
\centering
\includegraphics{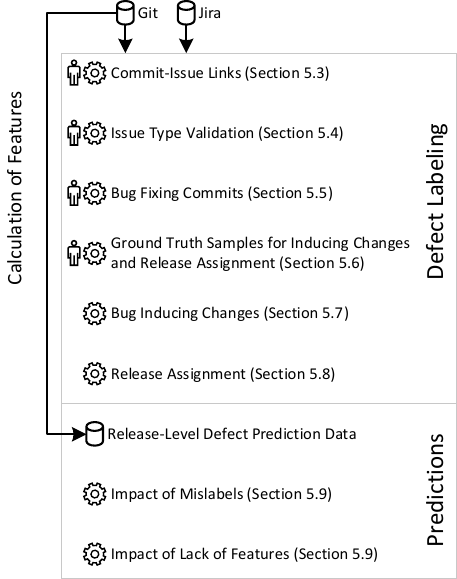}
\caption{Overview of the Experiment. The human figure indicates that we used manual validation.}
\label{fig:exp-overview}
\end{figure}

Within this section, we describe the results of an empirical study we conducted. The goal of this study was two-fold. On the one hand, we want to determine the impact of the problems we discussed in Section~\ref{sec:issues}. On the other hand, we want to determine if our proposed improvements can effectively resolve these problems. Figure~\ref{fig:exp-overview} outlines our experiment. First, we go step by step through the defect labeling. Then, we use the resulting defect prediction data to evaluate the effect of mislabels and the lack of features on the prediction performance of a defect prediction model.

For all of these analyses we only presented aggregated results for all projects we analyzed. The detailed tables with all data, including all code required for an exact replication of our work as well as the collected release-level and just-in-time defect prediction data sets, are part of the supplemental material.\footnote{https://doi.org/10.5281/zenodo.5675024}

\subsection{Data}
\label{sec:project_selection}

We conducted our study on projects from the Apache Software Foundation.\footnote{http://www.apache.org} Apache projects must have reached a certain level of maturity in order to be considered as a top-level project. Especially the use of Jira as issue tracking system is highly recommended and followed by most Apache projects. Additionally, \cite{Bissyande2013} found that ``Apache developers are meticulous in their efforts to insert bug references in the change logs of the commits''. This is an important property for the projects under consideration because it removes the need for link recovery and we can safely assume that links from commits to issues are available in the data. \cite{Yatish2019} also used a convenience sample of Apache projects for the same reason. A similar reasoning was also used for studies other than defect prediction, e.g., by \cite{DiPenta2020}. To further ensure the maturity of projects, we used the criteria listed in Table~\ref{tbl:project_selection_criteria}. In addition, we had a soft criterion that we focused on projects with less than 10.000 commits on the main development branch. The reason for this is the very high demand on the resources for the collection of the metric data for every commit in the Git repository.\footnote{While this criterion is irrelevant for the evaluation of the defect labeling, we selected the projects also with the goal to provide a new defect prediction data set.} Please note that the total number of commits can still be larger than 10.000 commits, because of we collect the data for all branches. 

\begin{table}
\centering
\begin{tabular}{lp{7.5cm}}
\textbf{Criterion} & \textbf{Rational} \\
\hline
Uses Git & Most projects either already use a Git repository, or provide a Git mirror of a SVN repository. \\
Java as main language & Our static analysis only works for Java code. \\
Uses Jira & The Jira of the Apache Software Foundation is the main resource for tracking issues of most Apache projects.\\
At least two years old & Project as a sufficient development history. \\
Not in incubator stage & Project has been fully accepted by the Apache Software Foundation. \\
$>$ 100 Issues in Jira & Project is mature and actively uses Jira \\
$>1000$ Commits & Project has a sufficient development history. \\
$>$ 100 Files & Project should have a reasonable size.\\
Activity since 2018-01-01 & Project is still active in both Jira and Git.\\
\hline
\end{tabular}
\vspace{1pt}
\caption{Criteria for the inclusion of projects.}
\label{tbl:project_selection_criteria}
\end{table}

Table \ref{tbl:projects} lists the 38 projects for which we collected data, including the versions of the 398 releases for which we collected release-level data. The releases were determined using the project homepages. For each release, we looked up the commit of the release in the Git repository. For most releases, there was a related tag in the Git repository. If this was not the case, we manually analyzed the commit history to determine the release commit, using the information we found on the project homepage, as well as related tags and branches.

\begin{table}
\centering
\scriptsize
\begin{tabular}{L{1.8cm}llL{6.2cm}}
\textbf{Project} & \textbf{Commits} & \textbf{Bugs} & \textbf{Releases} \\
\hline
ant-ivy & 3,189 & 535 & 1.4.1, 2.0.0, 2.1.0, 2.2.0, 2.3.0, 2.4.0  \\
archiva & 10,262 & 542 & 1.0, 1.1, 1.2, 1.3, 2.0.0, 2.1.0, 2.2.0  \\
calcite & 2,926 & 842 & 1.0.0, 1.1.0, 1.2.0, 1.3.0, 1.4.0, 1.5.0, 1.6.0, 1.7.0, 1.8.0, 1.9.0, 1.10.0, 1.11.0, 1.12.0, 1.13.0, 1.14.0, 1.15.0 \\
cayenne & 6,619 & 530 & 3.0.0, 3.1.0  \\
commons-bcel & 1,429 & 53 &  \textit{5.0}, \textit{5.1}, 5.2, 6.0, 6.1, 6.2 \\
commons-beanutils & 1,341 & 76 & \textit{1.0}, \textit{1.1}, \textit{1.2}, \textit{1.3}, \textit{1.4}, \textit{1.5}, \textit{1.6}, 1.7.0, 1.8.0, 1.9.0 \\
commons-codec & 1,838 & 64 &  \textit{1.1}, 1.2, 1.3, \textit{1.4}, 1.5, 1.6, 1.7, 1.8, 1.9, 1.10, \textit{1.11} \\
commons-collections & 3,380 & 115 &  1.0, 2.0, 2.1, 3.0, 3.1, 3.2, 3.3, 4.0, 4.1 \\
commons-compress & 2,755 & 172 &  1.0, 1.1, 1.2, 1.3, 1.4, 1.5, 1.6, 1.7, 1.8, 1.9, 1.10, 1.11, 1.12, 1.13, 1.14, 1.15, 1.16 \\
commons-configuration & 3,717 & 188 &  1.0, 1.1, 1.2, 1.3, 1.4, 1.5, 1.6, 1.7, 1.8, 1.9, 1.10, \textit{2.0}, \textit{2.1}, 2.2 \\
commons-dbcp & 2,205 & 127 &  1.0, 1.1, 1.2, 1.3, 1.4, 2.0, 2.1, 2.2.0, 2.3.0, 2.4.0, 2.5.0 \\
commons-digester & 2,525 & 26 &  \textit{1.0}, \textit{1.1}, \textit{1.2}, \textit{1.3}, 1.4, 1.5, 1.6, 1.7, 1.8, \textit{2.0}, \textit{2.1}, 3.0, 3.1, 3.2 \\
commons-io & 2,262 & 131 &  1.0, 1.1, 1.2, 1.3, 1.4, 2.0, 2.1, 2.2, 2.3, 2.4, 2.5 \\
commons-jcs & 1,622 & 80 &  1.0, 1.1, 1.3, 2.0, 2.1, 2.2 \\
commons-jexl & 3,276 & 84 &  \textit{1.0}, \textit{1.1}, 2.0, 2.1, \textit{3.0}, \textit{3.1} \\
commons-lang & 5,792 & 318 &  \textit{1.0}, 2.0, 2.1, 2.2, 2.3, 2.4, 2.5, 2.6, 3.0, 3.1, 3.2, 3.3, 3.4, 3.5, 3.6, 3.7 \\
commons-math & 7,222 & 415 &  1.0, 1.1, 1.2, 2.0, 2.1, 2.2, 3.0, 3.1, 3.2, 3.3, 3.4, 3.5, \textit{3.6} \\
commons-net & 2,270 & 176 &  1.0.0, 1.1.0, 1.2.0, 1.3.0, 1.4.0, 2.0, 2.1, 2.2, 3.0, 3.1, 3.2, 3.3, 3.4, 3.5, 3.6 \\
commons-scxml & 1,216 & 70 &  0.5, 0.6, 0.7, 0.8, 0.9 \\
commons-validator & 3,416 & 73 &  \textit{1.0}, 1.1.0, 1.2.0, 1.3.0, 1.4.0, 1.5.0, 1.6.0 \\
commons-vfs & 2,212 & 156 &  1.0, 2.0, 2.1, 2.2 \\
deltaspike & 2,311 & 302 &  0.1, 0.2, 0.3, 0.4, 0.5, 0.6, 0.7, 1.0.0, 1.1.0, 1.2.0, 1.3.0, 1.4.0, 1.5.0, 1.6.0, 1.7.0, 1.8.0 \\
eagle & 1,119 & 225 &  0.3.0, 0.4.0, \textit{0.5.0} \\
giraph & 1,121 & 337 &  \textit{0.1.0}, 1.0.0, 1.1.0 \\
gora & 1,329 & 113 &  0.1, 0.2, 0.3, 0.4, 0.5, 0.6, \textit{0.7}, \textit{0.8} \\
jspwiki & 8,809 & 274 &  1.4.0, 1.5.0, 1.6.0, 1.7.0, 1.8.0, 2.0.36, 2.2.19, 2.4.56, 2.6.0, 2.8.0, 2.9.0, 2.10.0 \\
knox & 2,069 & 568 &  0.3.0, 0.4.0, 0.5.0, 0.6.0, 0.7.0, 0.8.0, 0.9.0, 0.10.0, 0.11.0, 0.12.0, 0.13.0, 0.14.0, 1.0.0 \\
kylin & 12,975 & 732 &  0.6.1, 0.7.1, 1.0, 1.1, 1.2, 1.3, 1.5.0, 1.6.0, 2.0.0, 2.1.0, 2.2.0 \\
lens & 2,418 & 397 & 2.6.0, 2.7.0 \\
mahout & 4,167 & 513 &  0.1, 0.2, 0.3, 0.4, 0.5, 0.6, 0.7, 0.8, 0.9, 0.10.0, 0.11.0, 0.12.0, \textit{0.13.0} \\
manifoldcf & 5,936 & 633 &  0.1, 0.2, 0.3, 0.4, 0.5, 0.6, 1.0, 1.1, 1.2, 1.3, 1.4, 1.5, 1.6, 1.7, 1.8, 1.9, 1.10, 2.0, 2.1, 2.2, 2.3, 2.4, 2.5, 2.6, 2.7, 2.8, 2.9, 2.10 \\
nutch & 3,532 & 643 &  0.7, 0.8, 0.9, 1.0, 1.1, 1.2, 1.3, 1.4, 1.5, 1.6, 1.7, 1.8, 1.9, 1.10, 1.11, 1.12, 1.13, 1.14, 2.0, 2.1, 2.2, 2.3 \\
opennlp & 2,685 & 219 &  1.6.0, 1.7.0, 1.8.0 \\
parquet-mr & 2,249 & 1413 &  1.0.0, 1.1.0, 1.2.0, 1.3.0, 1.4.0, 1.5.0, 1.6.0, 1.7.0, 1.8.0, 1.9.0 \\
santuario-java & 3,376 & 83 &  \textit{1.0.0}, \textit{1.2}, 1.4.5, 1.5.9, 2.0.0, 2.1.0 \\
systemml & 6,196 & 452 & 0.9, 0.10, 0.11, 0.12, 0.13, 0.14, 0.15, 1.0.0, 1.1.0, \textit{1.2.0} \\
tika & 4,933 & 605 &  0.1, 0.2, 0.3, 0.4, 0.5, 0.6, 0.7, 0.8, 0.9, 0.10, 1.0, 1.1, 1.2, 1.3, 1.4, 1.5, 1.6, 1.7, 1.8, 1.9, 1.10, 1.11, 1.12, 1.13, 1.14, 1.15, 1.16, 1.17 \\
wss4j & 3,734 & 241 &  1.5.0, 1.6.0, 2.0.0, 2.1.0, \textit{2.2.0} \\
\hline
\end{tabular}
\vspace{1pt}
\caption{Apache projects and releases used for the empirical study. We did not assign any bugs to releases in italics as part of our empirical study.}
\label{tbl:projects}
\end{table}

We collected all data using the SmartSHARK~\citep{Trautsch2017, Trautsch2020} platform. The advantage of this approach is that we aggregate all collected data in a single MongoDB database, including the results of our manual validations. Details regarding the collection of this data can be found in Appendix~\ref{sec:collection-details}.

The data for just-in-time defect prediction is provided through the MongoDB. Additionally, we prepared release-level defect prediction data. As features we collected static code metrics, clone metrics, PMD warnings, AST node counts, the number of the different kinds of changes~\citep{Fluri2007} and refactorings~\citep{Silva2017} from the last six months, and churn metrics proposed by \cite{Moser2008}, \cite{Hassan2009}, and \cite{DAmbros2012}, and all thirteen aggregations that were proposed by \cite{Zhang2017} for the software metrics are not on the file level, i.e., class, method, interface, enum, attribute and annotation metrics. The data set contains a total of 4198 features.  We decided not to add features with code smells~\citep{Palomba2019} to the data set, because these can be calculated indirectly from the available source code metrics and definition of smells like god class may change over time. Additionally, features based on mutation testing are not available, because retrospective execution of tests is, unfortunately, often not possible~\citep{Tufano2017}. We also have not added features regarding test smells~\citep{Spadini2018} and review activity~\citep{Thongtanunam2016}, because current results only show correlation with defects, but have not yet shown that these features may actually improve defect prediction models. 

The defect labels are assigned using the JLMIV+IND approach. For each file, we stores the number of bugs that were fixed in this file. Moreover, the data contains a matrix that has as columns the issues and as rows the files. This matrix contains the value one, if the issue affected a file. This allows a fine-grained analysis which issues affected which file in a release and also which issues affected multiple files and is required for the analysis of costs~\citep{Herbold2019}. The column names of this matrix contain the identifier of the issue, the severity of the issue, and the date of the last bug fixing commit for the issue. This meta data allows for later filtering, e.g., to exclude trivial bugs or to ensure that there is no data leakage, e.g., to exclude bugs that were not yet fixed at the time of a release for which a prediction model is trained. 

\subsection{Evaluation Criteria}

For the evaluation of the different aspects of defect labeling, we determine baselines and then determine how well other approaches perform with respect to the baseline. The concrete baselines are discussed at the beginning of sections \ref{sec:issue-links}, \ref{sec:issue-types}, \ref{sec:bugfixes}, \ref{sec:inducing-changes}, and \ref{sec:release-assignments}. We evaluate two aspects: how many artifacts determined by the baseline are correctly identified and how many additional artifacts are identified. Artifacts are, for example, links from commits to issues, commits, or files. 
This approach is similar to the concept of true positives and false positives. For example, if a baseline determines $n$ artifacts as defective, and an approach for comparison A identifies $n_{tp}$ of these artifacts as defective as well, but also $n_{fp}$ additional artifacts, we say that A identifies $\frac{n_{tp}}{n}$ artifacts correctly as true positives and $\frac{n_{fn}}{n}$ additional false positive artifacts. For these comparisons, we report the median and the Median Absolute Difference (MAD) which is defined as $\text{MAD}=1.4826 \cdot median(|x_i-median(X)|)$ for a sample $X = \{x_1, ..., x_m\}$~\citep{Rousseeuw1993}. The median and the MAD have the advantage over the mean value and the standard deviation that they are robust in case of non-normal distributions, which is the case for most data in our empirical study. The value 1.4826 is a scaling factor for MAD that makes the values of MAD similar to the standard deviation of normally distributed data~\citep{Rousseeuw1993}. 

We use two independent researchers for the labeling of issue types of bug issues. We report Cohen's $\kappa$~\citep{Cohen1960} to estimate the reliability of the consensus, which is defined as
\begin{equation}
\kappa = \frac{p_0-p_e}{1-p_e}
\end{equation}
where $p_0$ is the observed agreement between the two raters and $p_e$ the probability of random agreements. $p_e$ depends on the number of categories $k$ and is defined as 
\begin{equation}
p_e = \frac{1}{N^2}\sum_{i=1}^k n_k^1\cdot n_k^2
\end{equation}
where $N$ is number of issues and $n_i^1$, resp. $n_i^2$ is the number of times researcher 1, resp. 2 labeled an issue as category $i$. We use the table from \cite{Landis1977} for the interpretation of $\kappa$ (see Table~\ref{tbl:kappa}).

\begin{table}
\centering
\begin{tabular}{cl}
$\kappa$ & \textbf{Interpretation} \\
\hline \hline
$<$0 & Poor agreement \\
0.01 – 0.20 & Slight agreement \\
0.21 – 0.40 & Fair agreement \\
0.41 – 0.60 & Moderate agreement \\
0.61 – 0.80 & Substantial agreement \\
0.81 – 1.00 & Almost perfect agreement \\
\end{tabular}
\caption{Interpretation of Cohen's $\kappa$ according to \cite{Landis1977}.}
\label{tbl:kappa}
\end{table}

We also evaluate defect prediction models as part of our empirical study, to evaluate the impact of mislabels and feature sets. To evaluate the practical relevance of differences between the models, we evaluate how training on noisy data or fewer features impacts the cost effectiveness of defect prediction models based on the cost model by \cite{Herbold2019}. The advantage of using the cost model by Herbold instead of commonly used metrics like recall, precision, or F-measure is that the values of these metrics are not reliable measures with respect to the cost saving potential of defect prediction models~\cite{Herbold2019}. Moreover, \cite{Yao2020} found that many commonly used measures are not able to reliably determine if models are better than random predictions and suggest to use Matthews Correlation Coefficient. However, since the relationship between MCC and costs is also unclear, which is why we decided to directly evaluate costs.

The cost model estimates boundaries on the ratio between costs of quality assurance and costs of defects. Defect predictions can save costs for a project based on the ratio $C$ between the costs for quality assurance and the cost of defects. The cost model estimates lower and upper boundaries for $C$ that must be fulfilled to allow the defect prediction model to save costs. For cost ratios $C$ less than the lower boundary, it would be better to not perform any additional quality assurance, because the quality assurance would be more expensive than the cost of the defects. The lower boundary increases with more predictions of files as defective (regardless whether the predictions are correct) and decreases with more defects being predicted. Thus, the lower boundary penalizes false positive predictions and rewards true positive predictions. For cost ratios $C$ greater than the upper boundary, it would be better to apply additional quality assurance everywhere, because the costs for the defects would be higher than the investment for the quality assurance. The upper boundary increases with fewer predictions of files as defects (regardless whether the predictions are correct) and decreases with more defects missed by the predictions. Thus, the upper boundary penalizes false negatives and rewards true negatives. An additional caveat of the cost model is that it accounts for the $m$-to-$n$ relationship between files and defects, i.e., that each file may be affected by multiple defects and the each defect may affect multiple files. As a result, the cost model does not use the confusion matrix for the estimation of true predictions of defects, but rather the bug-issue matrix to evaluate if all files that are affected by a defect are predicted by the defect prediction model. We use the size in logical lines of code as proxy for the effort for the quality assurance of the artifacts. Moreover, we assume that quality assurance is perfect, i.e, that all predicted defects are actually found by subsequent quality assurance measures. Following \cite{Herbold2019}, we can thus compute the boundaries on $C$ as
\begin{equation}
\frac{\sum_{s \in S: h(s)=1} size(s)}{|D_{PRED}|}
< C <
\frac{\sum_{s \in S: h(s)=0} size(s)}{|D_{MISS}|}
\end{equation}
$S$ is the set of files for which defects are predicted, $h$ is the defect prediction model, $D$ is the set of defects all $d \in \mathcal{P}(S)$\footnote{In the cost model, a defect is defined by the set of files that it affects.}, $D_{PRED} = \{d \in D: \forall s \in d~|~h(s) = 1\}$ is the set of predicted defects, and $D_{MISS} = \{d \in D: \exists s \in d~|~h(s)=0\}$ is the set of missed defects. 

We use Demsar's guidelines~\citep{Demsar2006} for the comparison of classifiers to evaluate if differences of the lower boundary and upper boundary are significant. The cost boundaries are not normally distributed and can even be infinite.\footnote{For example, the upper bound is infinite if no defects are missed and a single file is not predicted as defective.} Therefore, we use the Friedman test~\citep{Friedman1940} with the post-hoc Nemenyi test~\citep{Nemenyi1963} to evaluate significant differences. In case differences are significant, we use Cliff's $\delta$~\citep{Cliff1993} to estimate the effect sizes. According to \cite{Romano2006}, the effect is negligible if $\delta<0.147$, small if $0.147 \leq \delta < 0.33$, medium if $0.33 \leq \delta < 0.474$ and large if $\delta \geq 0.474$
Since there is no guarantee that the defect prediction model can actually save costs, i.e., that the lower boundary of the model is less than the upper boundary, we also report the percentage of releases for which the defect prediction cannot save costs because the lower boundary is greater or equal to the upper boundary. 



\subsection{Identification of Issue Links}
\label{sec:issue-links}

The first step of defect labeling is the identification of links between commits and issues. We restricted the analysis only to links to issues of type BUG that were closed and fixed at least once in their lifetime. Thus, we restrict this analysis to the links to issues that are relevant for defect labels. 

Our JLM approach that is based on a semi-automated validation of the links found by SZZ and JL identified 18,721 correct links from commits to issues. 5,311 of these links were manually validated by the first author of this article, the remaining 13,410 links were validated by our heuristic, i.e., had a link to a single issue directly at the beginning. We sampled 1000 of the links that were validated by our heuristic to evaluate the correctness of the heuristic. The heuristic was correct in all cases. Moreover, we randomly sampled 1000 commits from the all commits for which neither SZZ nor JL found a link to a Jira issue. We found no links to Jira that we failed to identify. However, we found 40 links to the Bugzilla\footnote{https://bz.apache.org} of the Apache Software Foundation. Since the data is not available anymore, we could not validate, if these issues are bugs or improvements. Therefore, about 4.0\% of the commits for which we found no link may also be bug fixing, but cannot be determined as such anymore because the issue data is missing. Otherwise, we found no errors in our data. Thus, while we believe that there may still be missed or invalid links in the data, the amount of data that is affected would be very small. Therefore, we can consider JLM as ground truth for the links between commits and Jira issues. We note that these findings are in line the empirical study by \cite{Bissyande2013} on issue links in Apache projects. 

\begin{figure}
\centering
\includegraphics[width=0.5\linewidth]{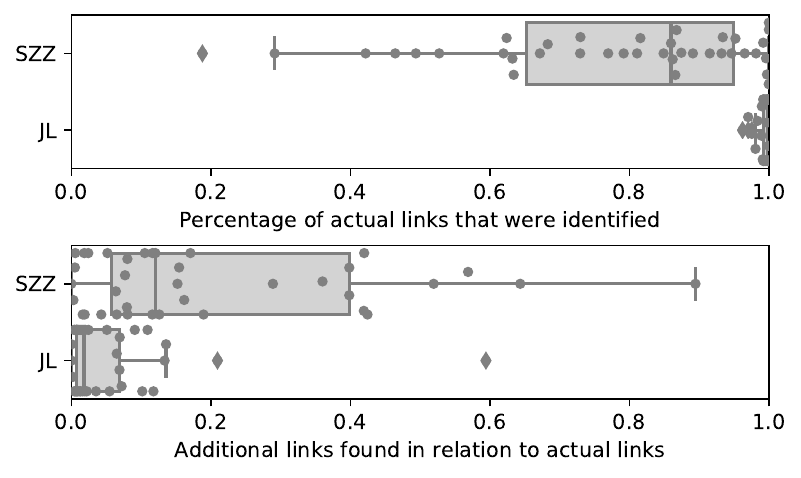}
\caption{Results of the validation of the link correctness. For the additional links with SZZ, outliers were cut of which were located at 124.3\% and at 430.4\%.}
\label{fig:links}
\end{figure}

We evaluate the performance of SZZ and JL with respect to the ground truth data determined with JLM. Figure~\ref{fig:links} summarizes the results. JL finds almost all correct links with a median of 99.7\% (MAD=0.4\%). In the worst case, JL still identifies 96.2\% of all links. However, JL finds a median of 1.7\% (MAD=1.9\%) additional links that are wrong. In the worst case, JL identifies 59.4\% additional links. This happened for the commons-jcs project and was due to the very frequent usage of version numbers within commit messages. Further investigation revealed that the usage of version numbers was the main reason for the false positive links by JL for all projects. 

SZZ finds a median of 85.4\% (MAD=19.6\%) of the correct links to issues. We note that the results of SZZ strongly vary, in the worst case SZZ only finds 18.8\% of the correct links. This happened for the commons-collections project. When we evaluated this, we found that for commons-collections, many issues were never assigned to a developer in the Jira. This breaks the semantic check of SZZ for the equality of the assignee in the issue tracker and the author of the commit. Further investigation revealed that this semantic check did not hold for all missed links by SZZ. Moreover, SZZ identifies a median of 12.3\% (MAD=15.5\%) additional links that are wrong. We note that while the median is relatively low, there is a long tail of projects with many additional links. There are even two outliers which are not shown in Figure~\ref{fig:links}. The outliers are for parquet-mr (430.4\% additional links) and cayenne (124.3\% additional links). In both cases, the broken links are due to links to pull requests on Github, which have the pattern \#$<$NUMBER$>$. Since SZZ cannot distinguish between different issue tracking systems, all these numbers are checked against the Jira of the projects and lead to additional links. Further investigation revealed that links to pull requests were the most frequent reason for additional links in general. Commonly used numbers were also problematic, but not as frequent. 

\subsection{Issue Type Validation}
\label{sec:issue-types}

The second part of the validation of the quality of the issue data is a partial conceptual replication~\citep{Shull2008} of the results by \cite{Herzig2013}. Based on the data for five projects by \cite{Herzig2013}, we expect that between 27.4\% and 42.9\% of BUG issues are mislabeled with 99.5\% confidence and that between 0.4\% and 3.5\% of issues of other types than BUG should actually be bugs. The first and third author of this article manually labeled the types for all linked issues of type BUG, regardless of whether the link was established by SZZ, JL, or JLM. The inter-rater reliability between the first and the third author was substantial ($\kappa=0.62$) with an agreement on 77\% of the issues. All three authors determined the correct label as committee for the remaining 23\% of the issues. This way, we manually validated the issue type for all 11,295 issues that were linked by commits and labelled as BUG in the issue tracker. 

\begin{figure}
\centering
\includegraphics[width=0.5\linewidth]{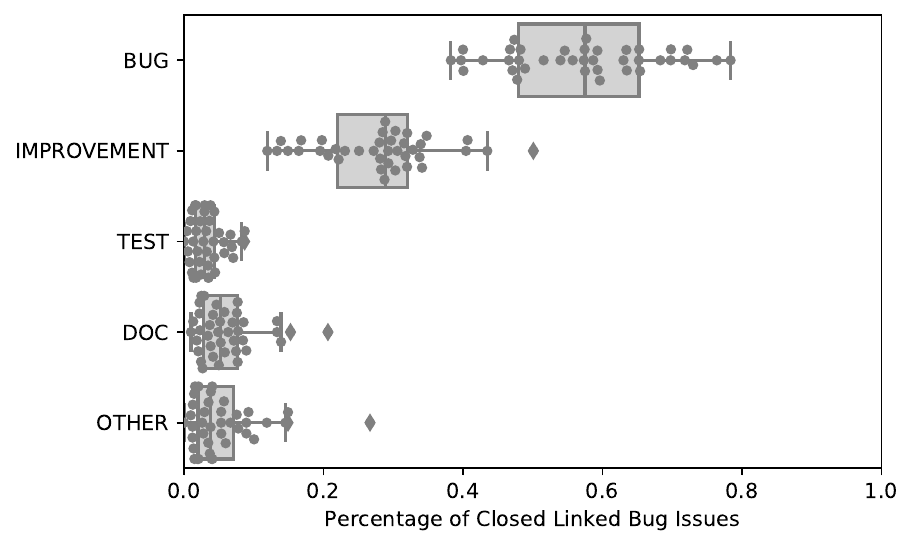}
\caption{Results of the manual validation of linked bug issues.}
\label{fig:issuetypes}
\end{figure}

Figure~\ref{fig:issuetypes} summarizes the results of the evaluation. Overall, we found that a median of 42.3\% (MAD=13.3\%) of linked BUG issues are mislabeled: 29.0\% (MAD=6.4\%) are actually IMPROVEMENT, 5.2\% (MAD=3.6\%) are DOC, 3.0\% (MAD=2.0\%) are TEST, and 3.8\% (MAD=3.1\%) are OTHER.\footnote{The sum of the median values for for IMPROVEMENT, DOC, TEST, and OTHER is 41.0\%, i.e., less than the median of not being a BUG, which is 42.5\%. This is possible because the median is not linear.} Thus, our results replicate the findings by \cite{Herzig2013} regarding the misclassification of BUG issues, even though we are close to the upper bound of the confidence interval. Another way to read these numbers is that for every correctly labeled BUG issue, there are a median of $\frac{42.3\%}{57.7\%}=0.73$ incorrectly labeled issues. Figure~\ref{fig:issuetypes} demonstrates that all projects are affected by this kind of noise in the data, i.e., even in the best case about one fifth of the bugs are mislabeled.

We also randomly sampled 50 issues that are not of type BUG for each project, i.e., a total of 1833 issues.\footnote{The projects commons-bcel, commons-digester, commons-jcs, and commons-validator had fewer than 50 issues that were linked by commits, which is way we do not have 38*50=1900 issues.} These issues were manually validated by the first and second author of this article and the first three authors resolved conflicts as committee. A mean of 1.3\% of these issues were actually bugs. Thus, our results also replicate these findings by \cite{Herzig2013}. Overall, there are 11980 resolved linked issues that are not of type bug in our data. Thus, we expect that we miss about $1.3\%\cdot11980 = 155.7$ bugs in our data. Since we have found 6367 validated bugs, we are missing about 2.4\% of the bugs in the data, i.e., the impact on our results is small.

\subsection{Bug Fixing Commits}
\label{sec:bugfixes}

\begin{figure}
\centering
\includegraphics[width=0.5\linewidth]{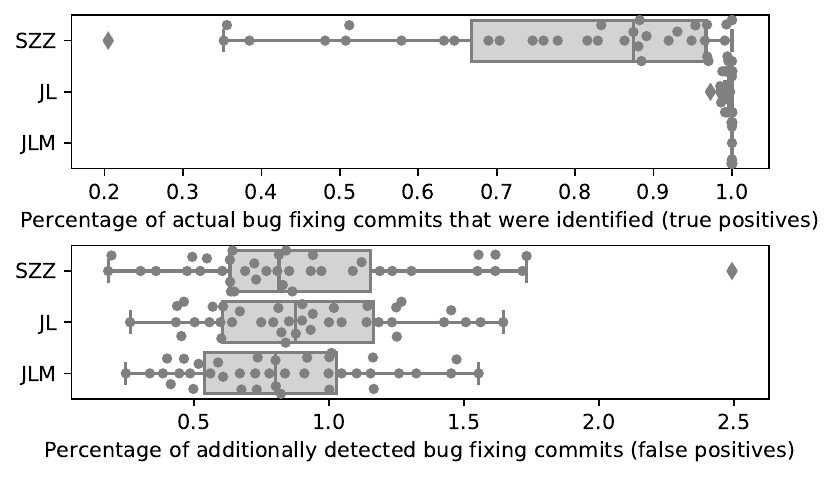}
\caption{Results of the validation of the link correctness.}
\label{fig:bugfixes-details}
\end{figure}

Neither the broken links, nor the wrong issue types have direct impact on defect prediction. The impact on defect prediction research only manifests, if there are false positives or false negatives for the labeling of bug fixing commits based on this data. To validate how the bug fixing commits change, we compare SZZ, JL, JLM, and JLMIV with each other, using JLMIV as baseline. JLMIV constitutes our ground truth for this evaluation, because it is based on the validated links and the validated issues. JLMIV labels a median of 5.6\% (MAD=3.3\%) of the total commits of a project as bug fixing. Figure~\ref{fig:bugfixes-details} summarizes the actual bug fixes that are detected correctly (true positives) and the wrongly detected bug fixing commits (false positives). The results for the true positives mirror the results of the detection between the commits and issues. SZZ identifies a median of 86.9\% (MAD=18.2\%) of all bug fixing commits, JL is almost perfect with a median of 100\% (MAD=0.4\%). JLM identifies all bug fixes identified by JLMIV, because JLMIV uses the same links as JLM and only reduces the bug fixing commits, because fewer issues are considered as BUG.  

Regarding the false positives, our results are mostly influenced by the issue type validation. SZZ finds a median of 81.1\% (MAD=40.0\%) false positive bug fixing commits, JL finds a median of 86.3\% (MAD=40.4\%), and JLM finds a median of 78.9\% (MAD=39.3\%). While these numbers are very high, they are expected given that there are a median of 0.73 wrong BUG issues for every correct bug issue. There are also additional false positives for SZZ and JL due to additional issue links that are wrong. We note that SZZ has a lower median than JL. This is counter intuitive, because SZZ should have more false positives, because SZZ has more additional issue links than JL. However, this is offset by the correct links that SZZ missed. These not only cause SZZ to miss true positives, but also to miss false positives due to wrong issue labels.

\subsection{Ground Truth for Inducing Changes and Affected Releases}
\label{sec:groundtruth-inducing}

We were able to establish ground truth data for all our empirical data so far. Unfortunately, this was not possible for us for the bug inducing changes and the assignment of bugs to releases, including the quality of the data in the affected versions field of Jira. When we considered if we could achieve this through manual validation, we came to the conclusion that this is impossible on this scale for a group of researchers (Section~\ref{sec:open-issues}). Even on a smaller scale, we could often not be sure if our assessment is correct, because we lack understanding of the details of the source code within the projects. We could also not re-use an established approach from the literature. \cite{Mills2018} determined similar ground truth data, but only for the file actions addressed in bug fixing commits, not their inducing counterparts. The projects that are part of our empirical study that were also investigated by \cite{Mills2018} are mahout and tika. \cite{Mills2018} sampled 34 issues for mahout and 22 issues for tika, 23 of these issues are validated bug fixes in our data. We cross-checked our data with these 23 issues to determine if all files that were found to be part of the bug fix would be detected as such by our approach. This is the case, if there is an inducing change for the file. We found that we correctly detect which file actions were actually part of the fix for 21 of the issues. For the issue TIKA-1110, we fail to identify one file, that \cite{Mills2018} say is part of the fix. However, this is the deletion of a file, because there are no further references to it. Since the deletion does not cause any change to the logic of the code, this is not a problem of our data. The cases where we miss changes are for TIKA-1070 and TIKA-961. In both cases, there is a pure addition of source code. Since pure additions do not have inducing changes, we cannot identify an inducing commit and would, therefore, miss the related file. Overall, our approach was correct for 21 of 23 issues, i.e., 91\% of the file actions. 

For the bug inducing changes, only \cite{Perez2020} created manually validated data regarding bug inducing changes. However, they target different projects than our work. To provide an indication about the inducing changes in our work, we wanted to rely on developers for this task. Thus, we looked for information in our data, which we could use to accurately determine inducing changes. We then compared the results to the work by \cite{Perez2020}.

Moreover, we require an approach different from the suggested framework for evaluating SZZ variants suggested by \cite{daCosta2017}. In their work, they suggest to use indicators for potential mislabels such as the disagreement with the affected versions field, the number of subsequent bug fixes, or the age of the bug, to identify potential mislabels. However, such anomalies only indicate that something may be wrong, but are not definitive indicators of mislabels, i.e., ground truth. Consequently, their results cannot be directly compared to our work.

We were able to extract knowledge about inducing changes from our data by exploiting the false positive links of our JL approach. One reason for false positives is that an issue is referenced in a commit message as the cause of a bug. Thus, we scanned all bug fixing commits identified by JLMIV for false positive issue links. We manually checked the commit messages and found twenty commits which clearly state that one issue was the cause of another issue and that also indicate that this is not just re-opening of the same issue again. For example, we found the following commit message: ``FIX: ChainResolverTest failures (IVY-882): The problem was due to the changes introduced in IVY-857. [...]''\footnote{https://github.com/apache/ant-ivy/commit/6d6d34} Thus, we can say that IVY-857 is the \textit{inducing issue} of IVY-882. It follows that the bug was in the software, when the work on the inducing issue was finished. Thus, we looked up the commits in our data, which were also linked to the inducing issue. In case there were multiple commits for the work on the inducing issue, we manually validated which of the commits on the inducing issue was the latest change that was related to the work on the fixed issue and marked this commit as the inducing commit. Then, we manually validated that these were indeed the commits that introduced the bugs. In all but one case, this was the latest commit on the inducing issue. The exception is the work on TIKA-2483,\footnote{https://github.com/apache/tika/commit/06486c} where the inducing change was not in the latest change,\footnote{https://github.com/apache/tika/commit/6930ff} but one of the prior changes.\footnote{https://github.com/apache/tika/commit/3aab15} We note that a similar approach was suggested in parallel work by \cite{Rosa2021}, who also looked for fixing commits where the message indicates when the bug was introduced. In their study, they found 129 such commits that referenced issues within a sample of 19,603,736 commits.

Table~\ref{tbl:inducingissues} lists the data we retrieved. We first note that ten of the issues we found only existed for less than five days in the projects. This is expected, because this data is not an unbiased sample from all bug fixing commits, but rather a sample of commits were developers identified a concrete issue as the reason. In these cases, a developer immediately noticed and fixed the problem and referenced the prior work. The other ten issues lived longer, one issue even existed for more than one year. We use this data to evaluate three aspects: 1) the accuracy of the detection of the latest inducing change with JLMIV+R; 2) the quality of the data in the affected versions field of the issue tracking system; and 3) the correctness of the assignment of the bugs to releases based on a six month timeframe (6M), the affected versions field (AV), and the inducing changes determined by JLMIV+ (IND). 

JLMIV+R finds the correct latest commit regarding the inducing file action for sixteen of the twenty issues. This is in line with the expectation from \cite{Perez2020}, who find that the correct commit is found 70\% of time, as well as \cite{Rosa2021} who also determined a precision of about 70\%. In four cases, JLMIV+R finds a commit that is newer than the actual inducing change. This is an expected weakness of JLMIV+R and of strategies that use the blame mechanism to find the most recent change in a version control system in general. This happens if there is a change on the defective source code between the inducing change and the bug fixing change. 

Regarding the affected versions field, we note that there are nine cases, in which the field was not used in the issue tracker. In seven of those cases, this is correct, as the bugs never affected a release, i.e., they were introduced and fixed between releases. In the other two cases, we validated that the bugs affected multiple releases of the software. Of the eleven cases, where the affected version field is defined, only one entry is completely correct (VALIDATOR-376). For two issues, the affected version fields contain a partially correct entry (TIKA-599 and TIKA-2483). In both of these cases, only the latest release is mentioned, the older releases which are also affected are ignored. The remaining eight entries of the affected version field are wrong, i.e., they list releases which are not affected by the bugs. In six cases, the bugs were fixed prior to the release (IVY-882, NUTCH-683, PARQUET-214, LENS-538, GIRAPH-88, GIRAPH-34), in the other two cases the bugs were only induced after the release (KYLIN-3223, EAGLE-573). In the first six cases, the developers assigned the version number of the release that is currently a work in progress, in the last two cases they assigned the version number of the latest release.

The problems with the affected version field are more severe than we expected. Overall, the bugs in this sample affect 10 releases, the affected version field only mentions three releases correctly. We expected that we would find this kind of error in the data of the affected version field, even though we expected fewer differences. Our analysis revealed that the affected versions field may also contain affected versions that are wrong, which we did not expect. The case were the version of the work in progress release is used, is relatively harmless for defect prediction: since the bug fixing commit is before the release, the commit will not be considered during the labeling of the release and the wrong value of the field will, consequently, be ignored. Similarly, our proposed improvement for the detection of inducing changes JLMIV+RAV would just use the date of the reporting of the bug and, therefore, also not be affected. The second case, where the latest release is entered, even though the bug was never in that release, is problematic. This leads to an additional assignment to a release and also breaks JLMIV+RAV as the actually inducing changes are after the date of this wrongly mentioned release and would, therefore, be flagged as suspect.

The last aspect is the assignment to releases. Assignment based on bug fixes six months after the release is correct for twelve issues, assignment based on affected versions is correct for eleven issues, and assignment based on the inducing changes for all twenty issues. The correctness of the six month time frame depends on two factors: the time to fix and the activity of the project. In case the time to fix was more than six months, the correct release was missed. In case the project was very active, e.g., with multiple releases within the last six months, the bug would be assigned to a release in which it was not yet introduced into the software. With the exception of VALIDATOR-273, the release assignment based on the affected version field is correct if no release was affected and the affected versions either contained a not yet released version or was empty. The assignment based on inducing changes is correct, even in the four cases where a wrong change is identified as inducing. In one of these cases, there is only a small deviation of less than one week between the actual time to fix and the determined inducing change. In case of PARQUET-214 and VALIDATOR-376 the inducing change is relatively far off and it is pure chance due to the project activity that the assignment is correct. 

\begin{table*}
\footnotesize
\begin{sideways}
\centering
\begin{tabular}{llrR{2cm}L{2cm}L{2cm}C{0.3cm}C{0.3cm}C{0.3cm}}
\textbf{Fixed Issue} & \textbf{Inducing Issue} & \textbf{Time to Fix} & \textbf{Deviation JLMIV+R} & \textbf{Affected Versions Field} & \textbf{Actually Affected Releases} & \textbf{6M} & \textbf{AV} & \textbf{IND} \\
\hline
IVY-882  &  IVY-857  &  45 days &  0 days &  2.0-RC1  & - & \faCheck & (\faCheck) & \faCheck \\
CALCITE-2253  &  CALCITE-2206  &  5 days &  0 days & - & - & & \faCheck & \faCheck \\
CALCITE-1215  &  CALCITE-1212  &  0 days &  0 days & - & - &  & \faCheck & \faCheck \\
CALCITE-822  &  CALCITE-783  &  17 days &  0 days & - & - &  & \faCheck & \faCheck \\
KYLIN-3223  &  KYLIN-3239  &  1 days &  0 days &  2.2.0  & - &  & & \faCheck \\
NUTCH-683  &  NUTCH-676  &  20 days &  0 days &  1.0.0  & - & \faCheck & (\faCheck) & \faCheck \\
PARQUET-214  &  PARQUET-139  &  54 days &  -29 days &  1.6.0 & - & \faCheck & (\faCheck) & \faCheck \\
TIKA-599  &  TIKA-528  &  147 days &  0 days &  0.9 & 0.8, 0.9 & \faCheck & & \faCheck \\
TIKA-2483  &  TIKA-2311  &  196 days &  0 days & 1.16 & 1.15, 1.16 & \faCheck & & \faCheck \\
SYSTEMML-1126  &  SYSTEMML-584   &  323 days &  -7 days & - & 0.10, 0.11, 0.12, 0.13 &  & & \faCheck \\
SYSTEMML-2162  &  SYSTEMML-1919  &  163 days &  0 days & - & 1.0.0 & & & \faCheck \\
SYSTEMML-2275  &  SYSTEMML-2217  &  23 days &  0 days & - & - & & \faCheck & \faCheck \\
LENS-538  &  LENS-486  &  0 days &  0 days &  2.2 & - & \faCheck & (\faCheck) & \faCheck\\
KNOX-1134  &  KNOX-1119  &  2 days &  0 days & - & - & & & \faCheck \\
VALIDATOR-376  &  VALIDATOR-273 &  474 days & -174 days & 1.4.1 & (1.4.1) & \faCheck & \faCheck & \faCheck \\
GIRAPH-918  &  GIRAPH-908 &  2 days &  0 days & - & - & \faCheck & & \faCheck \\
GIRAPH-832  &  GIRAPH-792 &  1 days &  0 days & - & - & \faCheck & & \faCheck \\
GIRAPH-88  &  GIRAPH-11 &  0 days &  0 days &  0.1.0 & - & \faCheck & (\faCheck) & \faCheck\\
GIRAPH-34  &  GIRAPH-27 &  4 days &  -1 days &  0.1.0, 1.0.0  & - & \faCheck & (\faCheck) & \faCheck \\
EAGLE-573  &  EAGLE-569 &  0 days &  0 days &  0.5.0  & - & \faCheck & & \faCheck \\
\hline
\end{tabular}
\end{sideways}
\caption{Ground truth data for inducing changes and release assignments. The affected version for VALIDATOR-376 is in braces because it is minor version, which we omitted otherwise.}
\label{tbl:inducingissues}
\end{table*}

\subsection{Bug Inducing Changes}
\label{sec:inducing-changes}

For just-in-time data, the identification of bug fixing commits is only the precursor for finding the inducing changes, which are then the target of the prediction. Moreover, we described how the inducing change can be used for assigning defects to releases in Section~\ref{sec:improvements}. To this aim, we compare four approaches for the identification of bug inducing changes: 1) the standard SZZ algorithm 2) JLMIV, i.e., our improved linking with the issue validation, but standard SZZ to determine inducing changes; 3) JLMIV+R, i.e., the improvement to ignore changes to non-java files, whitespace only changes, documentation changes, and refactorings; 4) JLMIV+AV that further extends JLMIV by using the affected versions field; and 5) SZZ-RA, i.e., a state of the art variant of SZZ based on the work by \cite{Neto2018, Neto2019}. Our variant of SZZ-RA differs from the original work only in details: we added a filter that ignores test and documentation files, and use the links between commits and issues based on the Jira issue pattern.


We do not have ground truth data for this analysis, as discussed in Section~\ref{sec:groundtruth-inducing}. Instead, we use JLMIV+R as a baseline and analyze the differences between SZZ, JLMIV, JLMIV+AV and SZZ-RA with respect to JLMIV+R. The reason for this is two-fold. First, the inducing changes of JLMIV+R are a subset of JLMIV that only reduces the inducing changes, e.g., due to whitespace changes. Thus, in case of deviations, the change identified by JLMIV is always a false positive. Second, JLMIV+R is based on our ground truth for bug fixing commits. Since SZZ uses the same inducing strategy as JLMIV, but is based on the inferior SZZ labels for bug fixing commits, all deviations of SZZ from JLMIV+ are also mislabels. Similarly, all deviations from SZZ-RA from JLMIV+R are mislabels, because of wrong bugfixing commits. Regarding JLMIV+AV, we cannot state whether deviations from JLMIV+ are correct or not: this depends on the affected version field. In case the affected version field contains valid data, JLMIV+AV is likely to be correct, because the identification of suspect changes is improved. In case of invalid data, JLMIV+R is likely to be correct, because the inducing changes would be wrongly flagged as suspect by JLMIV+AV.  

\begin{figure}
\centering
\includegraphics[width=0.6\linewidth]{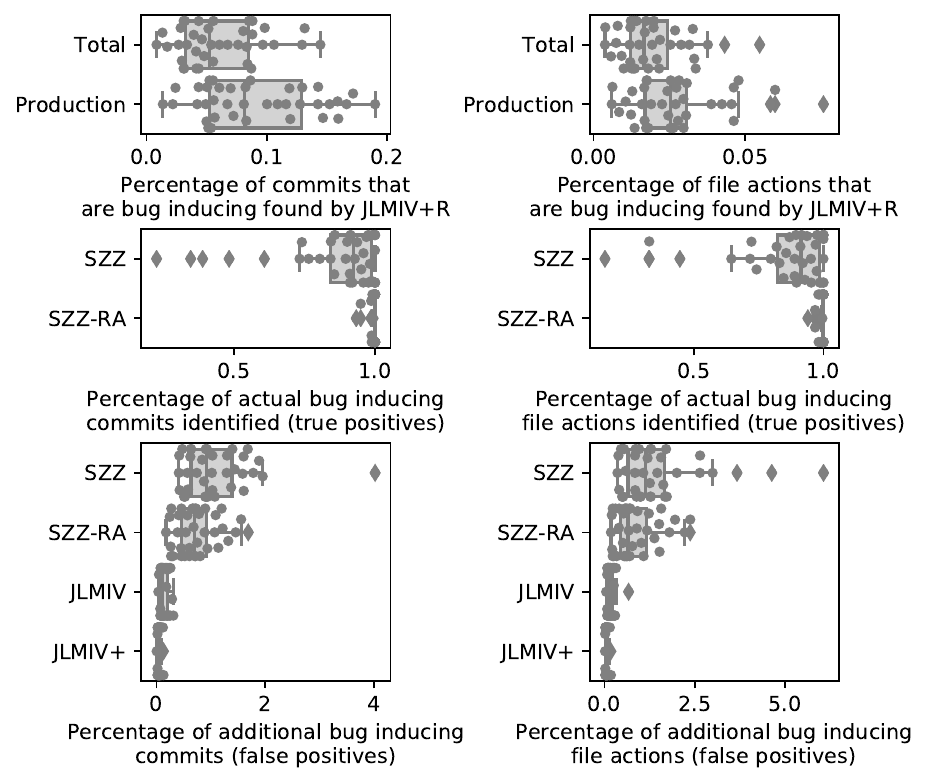}
\caption{Results for the identification of bug inducing changes.}
\label{fig:inducing-changes}
\end{figure}

Figure~\ref{fig:inducing-changes} summarizes the results for the inducing strategies. JLMIV+R finds that a median of 5.2\% (MAD=3.4\%) of the commits are bug inducing. If we only consider the 78.296 commits in which at least one Java production file\footnote{Java files excluding tests and documentation.} was changed, the percentage of bug inducing commits has a median of 8.1\% (MAD=5.0\%). When we consider this on the level of changes to files, as is done by \cite{Pascarella2019}, we find that a median of 2.5\% (MAD=1.2\%) of all changes to Java production files are inducing for a bug. 

SZZ identifies a median of 92.4\% (MAD=10.7\%) of the correct bug inducing commits and a median of 92.7\% (MAD=53.6\%) false positive bug inducing commits. SZZ identifies a median of 91.3\% (MAD=11.9\%) of the correct bug inducing file actions and a median of 113.3\% (MAD=80.6\%) of false positive inducing file actions of Java production files. These values are similar to the results for the bug fixing commits, i.e., mislabels due to the inducing strategy are hidden due to the large number of mislabeled bug fixes. The evaluation of JLMIV gives a better insight into the inducing strategy, because there is no noise due to mislabeled bug fixing commits. JLMIV identifies a median of 12.1\% (MAD=6.2\%) false positive bug inducing commits and a median of 13.3\% (MAD=6.7\%) false positive inducing file actions for java production files. This reduction is in line with the expectations due to the results from \cite{Mills2018}, which found that 8.7\% of false positives for the bug fixing actions are due to changes to comments and whitespace only changes and 8.49\% false positives are due to refactorings. 

With respect to JLMIV+AV, we find a median of 4.8\% (MAD=5.2\%) commits and a median of 4.9\% (MAD=4.2\%) of file actions are detected less than with JLMIV+R. Based on our limited data regarding the correctness of the affected versions field, we would expect that roughly half of these file actions are actually false positives (Section~\ref{sec:groundtruth-inducing}), i.e., are incorrectly detected by JLMIV+ and constitute noise. Thus, the potential impact of the affected versions field is relatively small with an expected reduction of false positive inducing changes by about 2.4\% of the total amount of inducing file actions. Regardless, we cannot recommend to use JLMIV+AV without first validating the data in the affected version field, because the potential benefit due to fewer false positives are offset by an equally large loss due to false negatives.

SZZ-RA similar as for the bugfixing commits, SZZ-RA finds almost all correct bug inducing commits with a median of 100\% (MAD=0.0\%) and a minimum of 93.3\%. However, SZZ-RA finds a median of 69.8\% (MAD=35.6\%) false positive bug inducing commits and a median of 65.0\% (MAD=48.7\%) of false positive inducing file actions. Thus, the percentage of false positive inducing commits is lower than the percentage of false positive bug fixing commits (86.3\%), but still relatively high. This reduction is in line with the expectation of the effect of the improvements to SZZ, i.e., ignoring changes to test, documentation, whitespaces, and refactorings, as can be seen by the difference between JLMIV and JLMIV+R. Thus, improved SZZ heuristics can resolve the relatively small problems of missing bug inducing changes due to missed issue links, but cannot resolve the problem with false positives due to wrong issue types. 

\subsection{Assignment to Releases}
\label{sec:release-assignments}

The literature suggests either to assign all bugs that are fixed within six months after a release to the release (6M) or to use the affected versions field (AV). In this article, we propose to use the inducing changes instead (IND). We evaluate the release assignment from two perspectives: the assigned issues and the files that are labeled as defective, due to the assigned issues. Same as for the inducing changes, we do not have ground truth. Regardless, the results from Section~\ref{sec:groundtruth-inducing} indicate that the assignment based on IND is the most reliable strategy, even though likely not flawless. Therefore, we evaluate the deviations of 6M and AV from IND. We use the 6M strategy with the bug fixing commits determined by SZZ (6M-SZZ), SZZ-RA\footnote{Technically, we only applied SZZ-RA for inducing changes. By considering which bug fixes identified by JL have an inducing change identified by SZZ-RA, we applied the improvements of SZZ-RA over SZZ to the bug fixing commits for the release assignment.} (6M-SZZ-RA), as well as those determined by JLMIV (6M-JLMIV). For the affected versions, we also use the bug fixing commits determined by JLMIV (AV-JLMIV). For the the assignment based on incuding changes, we use JLMIV+R (IND-JLMIV+R).

\begin{figure}
\centering
\includegraphics[width=0.6\linewidth]{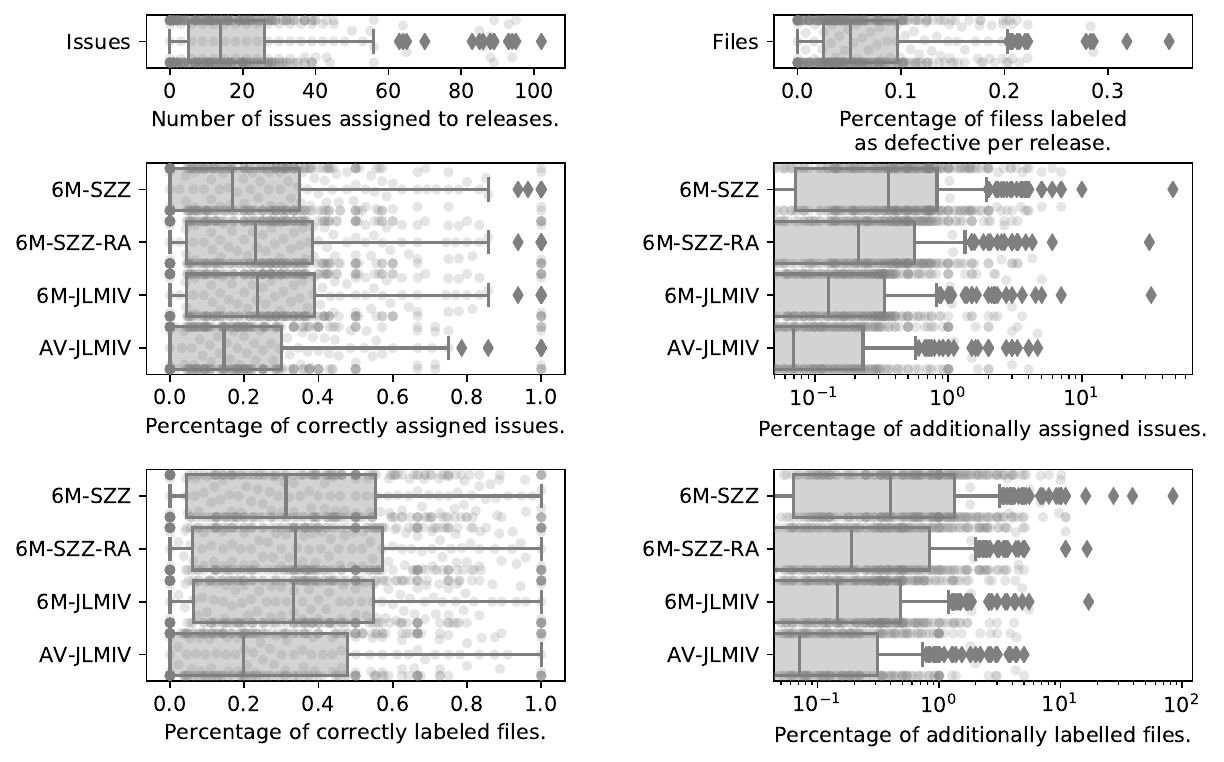}
\caption{Results for the assignment of issues to releases and the labeling of files.}
\label{fig:releasedata}
\end{figure}

Figure~\ref{fig:releasedata} summarizes the results for the release assignments. Overall, a median of 14 (MAD=14.8\%) bugs affect a median of 13.0\% (MAD=14.8\%) files of a release. For 36 releases, we did not find any bug fixes. We marked these releases in italic in Table~\ref{tbl:projects}. 21 of these release are the first releases for the projects, ten releases are the last in our data. The other five releases are for stable versions of Apache Commons projects. Without these 36 releases, the median of bugs that affect a release is 16 (MAD=13.3\%) and 15.5\% (MAD=14.1\%) files are defective. We report the results of 6M-SZZ, 6M-SZZ-RA, 6M-JLMIV and AV-JLMIV without the 36 releases that do not have any bug fix as they may skew the results. 

6M-SZZ determines a median of 16.9\% (MAD=25.1\%) of the bugs and a median of 33.3\% (MAD=38.5\%) of the files correctly, and determines 33.3\% (MAD=49.4\%) additional bugs and 37.8\% (MAD=56.0\%) false positive defective files. 6M-SZZ-RA determines a median of 23.1\% (MAD=26.5\%) of the bugs and a median of 33.8\% (MAD=38.7\%) of the files correctly, and determines 20.0\% (MAD=29.6\%) additional bugs and 17.0\% (MAD=25.2\%) false positive defective files. 6M-JLMIV determines a median of 23.6\% (MAD=26.7\%) of the bugs and a median of 33.3\% (MAD=35.3\%) of the files determined correctly and a median of 11.9\% (MAD=17.7\%) additional bugs and 14.3\% (MAD=21.2\%) false positive defective files. The differences between 6M-SZZ, 6M-SZZ-RA and 6M-JLMIV are in line with the differences in the inducing changes. AV-JLMIV determines a median of 14.5\% (MAD=21.6\%) of the bugs and a median of 19.8\% (MAD=29.4\%) of the files correctly and determines 6.4\% (MAD=9.6\%) additional bugs and 6.6\% (MAD=9.7\%) false positive detective files. Thus, AV-JLMIV labels the fewest files as defective of all variants. This is in line with the results by \cite{daCosta2017} that only a small percentage of bugs contain any data in the affected versions field.
We compared these results with the ground truth data from Section~\ref{sec:groundtruth-inducing}. While the deviations are not equal, they show similar trends to the sample depicted in Table~\ref{tbl:inducingissues}, both regarding the 6M strategy, as well as the AV strategy. 

\subsection{Prediction Models}
\label{sec:predictions}

The focus of our empirical analysis is on the quality of defect labels. However, it is unclear if and how the mislabels affect prediction models. Moreover, we have no indication if the incomplete feature sets in many data is really a problem. To assess the impact our data collection may have on the defect prediction models, we evaluate how the assessment of the best performing model from a recent benchmark on cross-project defect prediction~\citep{Herbold2017} would change. We follow the procedures for data processing outlined by \cite{Herbold2017} and use only the releases with at least 100 files and at least five defective files. We apply this criterion to both the 6M-SZZ and the IND-JLMIV+R data, i.e., there must be at least five defective files with both labels. We then use all data from projects other than the release for training, the data from the release itself for testing. This leaves us with 203 releases. \cite{CamargoCruz2009} proposed a relatively simple transfer learning approach. The approach by \cite{CamargoCruz2009} ranked best overall among 450 different models from the defect prediction literature that were compared on five data sets by \cite{Herbold2017} and should, therefore, be a suitable model to assess the impact of mislabels and feature sets on defect prediction models. 

Carmargo Cruz and Ochimizu proposed the following: first, the natural logarithm is applied to all features values plus one, then the difference between the median of the feature in the training data and the test data is subtracted, i.e., the feature $m$ is transformed such that $\hat{m}(s) = \log
(1+m(s))+median(\log(1+m(S)))-median(\log(1+m(S^*)))$ where $S$ is the set of files in the training data, $S^*$ is the set of file in the test data, $m(s)$ is the value of feature $m$ for file $s \in S \cup S^*$. The transformed features are used as input into a Gaussian Naive Bayes classifier~\citep{Zhang2004}. 

To evaluate the impact of features and labels on the prediction model, we train four different variants of the model by \cite{CamargoCruz2009} based on different training data. We use two different labels: 6M-SZZ, i.e., the commonly used labeling strategy used by most defect prediction data sets and IND-JLMIV+R, i.e., the most accurate labels we determined. For each label, we use two different sets of features: 1) all features in our data (ALL) and 2) only features based on static product metric for classes\footnote{We use summation for the aggregation to the file level.} and files (SM), i.e., the similar features to the PROMISE data, one of the most popular defect prediction data sets~\citep{Hosseini2017}. We denote the resulting models in the following as 6M-SZZ-ALL, 6M-SZZ-SM, IND-JLMIV+R-ALL, and IND-JLMIV+R-SM. These models are evaluated on the IND-JLMIV+R labels in the test data, i.e., with the least amount of mislabels. Through these models, we evaluate how mislabels affect the training of prediction models. Additionally, we also evaluate the models trained with 6M-SZZ labels with 6M-SZZ labels as test data and denote these models as 6M-SZZ-ALL-SZZ and 6M-SZZ-SM-SZZ. Through these models, we evaluate how mislabels affect the evaluation of prediction models. 

\begin{figure}
\centering
\includegraphics[width=\linewidth]{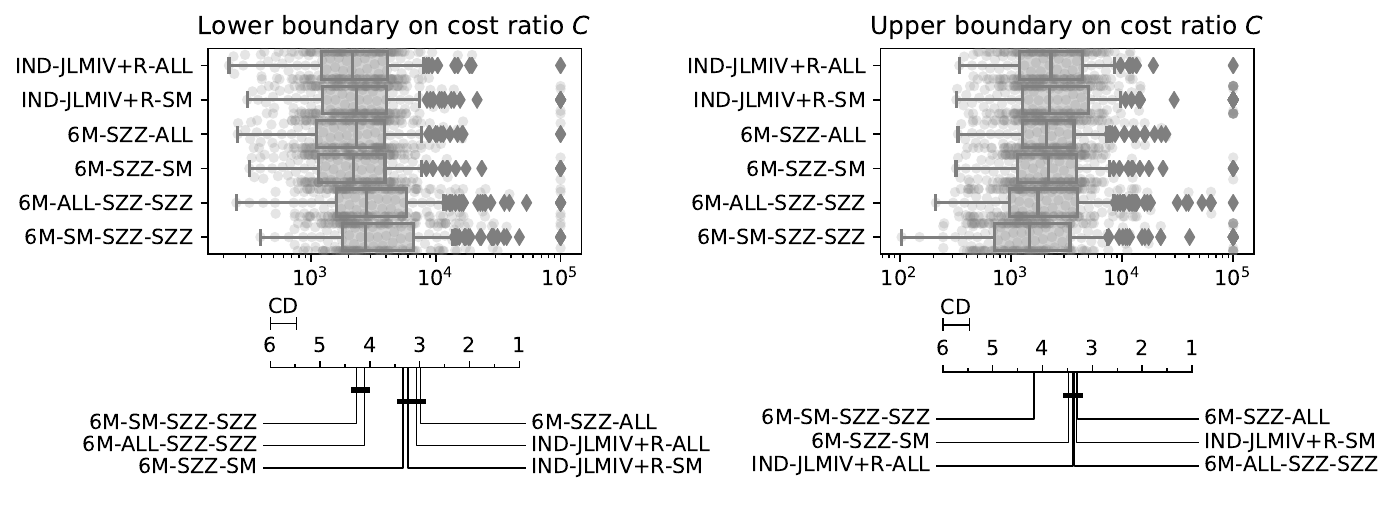}
\caption{Critical Distance diagrams for the comparison of significant differences the boundaries on the cost ratio $C$}
\label{fig:predictions}
\end{figure}

Figure~\ref{fig:predictions} summarizes the results of the prediction models. For IND-JLMIV+R-ALL the median lower boundary on $C$ is 2141.3 (MAD=1804.9) and the median upper boundary on $C$ is 2297.3 (MAD=1993.0). Thus, in the median, the costs of a defect must be at least as high as the quality assurance for 2141.3 lines of code and at most as high 2297.3 lines of code, for the model to be cost saving, i.e., the defect prediction model can only save cost under very specific assumptions in the median. For 86 of the 203 releases we use, the lower boundary is greater or equal to the upper boundary, meaning that for 42.4\% of the projects the defect prediction could never save costs in comparison to a trivial approach of either doing nothing or testing everything. While these numbers are already bad, they are worse for the other models. IND-JLMIV+R-SM has a median lower boundary of 2315.2 (MAD=1817.5) and a median upper boundary of 2221.5 (MAD=1827.6), the lower boundary is greater or equal to the upper boundary for 50.7\% of the projects. 6M-SZZ-ALL has a median lower boundary of 2313.4 (MAD=1996.8) and a median upper boundary of 2094.3 (MAD=1715.0), the lower boundary is greater or equal to the upper boundary for 46.3\% of the projects. 6M-SZZ-SM has a median lower boundary of 2185.0 (MAD=1754.3) and a median upper boundary of 2185.0 (MAD=1754.3), the lower boundary is greater or equal to the upper boundary for 45.3\% of the projects. When we train and test with the SZZ labels, the cost estimations get even worse. 6M-SZZ-ALL-SZZ has a median lower boundary of 2781.6 (MAD=2373.4) and a median upper boundary of 1755.1 (MAD=1652.8), the lower boundary is greater or equal to the upper boundary for 72.4\% of the projects. 6M-SZZ-SM-SZZ has a median lower boundary of 2724.8 (MAD=2067.9) and a median upper boundary of 1467.8 (MAD=1334.5), the lower boundary is greater or equal to the upper boundary for 78.3\% of the projects. Thus, for all models except IND-JLMIV+R-ALL the median lower boundary is greater or equal to the median upper boundary and the defect prediction model saves costs for fewer projects.

We reject the null hypothesis of the Friedman test that there are no significant differences between models for both the lower boundary (p-value$<$0.001) and the upper boundary (p-value$<$0.001). Figure~\ref{fig:predictions} shows the critical distance diagrams of the post-hoc Nemenyi test for both boundaries. The results show that the differences in the median between the models evaluated with IND-JLMIV+R labels in the test data are not significant. Thus, contrary to \cite{Tantithamthavorn2015}, we do not find that mislabels significantly impact the results, at least in terms of cost saving potential. The same is true for using more features. However, we note that while the difference may not be significant, using more features and cleaner labels still has the best median values and can be cost saving for the largest number of projects. 

The models tested with 6M-SZZ labels in the test data are significantly worse with a small effect size ($\delta \in [0.18, 0.23]$), with the SM features in both boundaries and with ALL features only in the lower boundary. This difference also manifests in a large drop in projects, for which the evaluation on the test data estimates that costs could be saved. 

Finally, we note that the overall performance of the model is relatively poor. However, the goal of this experiment is not to show whether the model by Carmargo Cruz and Ochimizu is good, but rather to evaluate if the different labels and features impact the evaluation of the model.

\section{Discussion}
\label{sec:discussion}

Within this section, we discuss the implications of the results of our empirical study on our research question, i.e., how issues with defect labeling and feature sets affect defect prediction data and prediction results. We consider three different aspects: the defect labels, the impact of features on predictions, the low performance of the defect prediction model we trained, as well as open problems. 

\subsection{Defect Labels}

Through our empirical study, we showed that the concerns that were raised in the recent years regarding the validity of defect labeling are valid. The two biggest factors that influence the quality of defect labels determined by SZZ are mislabeled issues in the Jira and mistakenly assigned bugs to releases. We proposed a solution for the latter problem by assigning bugs to releases based on inducing commits. The problem of mislabeled issues is more severe: to the best of our knowledge, there is no automated heuristic that can identify mislabeled issues in issue trackers. Our approach was to employ manual analysis by experts, same as \cite{Herzig2013}. However, this is extremely time consuming and does not scale well. In our case, we validated 11295 issues overall for the 38 projects. While we did not measure the exact time, two authors of this article spent at least one person month each on the independent labeling, i.e., at least 176 hours. Additionally, all three authors spent at least 20 hours on the resolution of disagreements. Thus, we spent at least 412 working hours on the issue type validation, the actual number is more likely around 600 working hours. Consequently, we required between two and three minutes per issue. This estimate is similar to the four minutes that \cite{Herzig2013} reportedly required. A third factor are wrong links from commits to issues, but this can already be automatically solved by adopting the linking process to the issue tracking system, in our case Jira. However, if this is not done, this would be a third major source for noise. We note that according to our analysis of existing defect prediction data sets (see Section~\ref{sec:related-work}), only RNALYTICA and BUGHUNTER data avoid this problem by exploiting links from Jira (RNALYTICA) and within GitHub (BUGHUNTER). This is an indication that the state of practice may be evolving here away from the original SZZ for the identification of bug fixing changes.

We note that the mislabeled issues are mislabeled from our perspective as researchers that want to analyze defects in software repositories. Thus, issues like incompatibilities due to new Java versions, failing tests, or missing documentation, are mislabels, because these do not constitute bugs in the sense of wrong run-time behavior of the software at the time of the release. From the perspective of the developers, these may not be mislabels, because they may use a more practical definition of bug in the issue tracker: something that is undesirable. Thus, we believe that these mislabels will remain a systematic problem for the analysis of issue tracking data, same as \cite{Herzig2013}.

Regarding the impact of the problems with defect labeling, our study revealed that all problems are replicable and that the impact increases if the problems are considered together. The problems with issue linking and the issue types in the repository combined mean that \textbf{SZZ misses about one fifth of the bug fixing commits, and only about half of the commits SZZ identifies as bug fixing are actually bug fixing}. This is because SZZ only identifies about 80\% of the actual bug fixing commits, but also mislabels roughly the same amount of bug fixing commits because of the mislabeled issues. If this is combined with a six month time frame for assigning defects to releases, the problem becomes even more severe. \textbf{For every file, that is correctly labeled as defective, there is roughly one file that is incorrectly labeled as defective, and two files that are incorrectly labeled as non-defective.} With implementation of the state of the art SZZ-RA variant of SZZ, these problems can be somewhat mitigated, but are still severe, because the main reasons for the deviates are the mislabeled issues and the six month time frame: \textbf{SZZ-RA misses almost no bug fixing commits but still has the problem that only about half of the commits identified as bug fixing are actually bug fixing. For every three files that are correctly labeled as defective, SZZ-RA incorrectly labels two files as defective, and misses six files that are incorrectly labeled as non-defective}. 
\cite{Yatish2019} proposed using the affected version field of issue trackers as a solution for the release assignment problem. While this is in principle a perfect solution, the reality of the data in the issue tracking system shows that the information contained in the affected version field is unreliable. Overall, using inducing changes currently seems to be the best heuristic. 

Another important aspect to consider here is, that there is still noise in our data due to false positive defect labels. We did not validate the file actions and instead only applied a heuristic that ignored changes to whitespaces, comments, and refactorings. Thus, while we may have measured the largest part of the noise in the defect labels, there is still an uncertain region, which is also not accounted for in our data. We note that additional improvements would only lead to more deviations of SZZ from the actual data, because SZZ would identify even more false positives. 

All data sets we discussed in Section~\ref{sec:related-work} are affected by the problems regarding the defect labels. This is a severe threat to all publications that use this data and, therefore, basically to the complete state of the art of defect prediction. We have shown in Section~\ref{sec:predictions} that the mislabels lead to significantly different results when used for training and evaluation. The differences are not significant if noisy data is used for training but cleaned data is used for testing. How this impacts, e.g., the comparison between defect prediction approaches is unclear. It may be that empirical results are not affected, because the signal of the defective data was still strong enough to be picked up by analysis and all approaches are affected equally. It may also be that the outcome of experiments changes, because different software artifacts are labeled as defective. 

\subsection{Features}

Our initial motivation that started this research was, that we actually wanted to have a defect prediction data set with a broad selection of features, because we believed that this was the key ingredient that was missing for highly performing defect prediction models. The indications from the literature suggested that this was true (Section~\ref{sec:lack-of-features}), especially due to churn related features. Because we considered how we should collect the defect data, we discovered the need for our analysis of the defect labels and the focus of our research evolved: due to our findings regarding the defect labels, the analysis of the features is now only in the background of this article. Regardless, we believed that the small experiment we described in Section~\ref{sec:predictions} would show that the large feature set is beneficial and that we could point to future research to find a suitable subset from the large amount of features we use, e.g., to minimize the effort to collect the features without a loss in prediction performance. 

Our results do not support such a conclusion and instead indicate that the impact of features that go beyond static source code metrics may be relatively small. While the results were best with all features, the difference was not significant. Our replication package also contains the results for \RECALL{} and \PRECISION, which show that the differences with traditional machine learning metrics are also not very strong, i.e., insignificant for \RECALL{} and potentially small for \PRECISION. A closer look at the literature reveals a potential reason for this lack of a stronger effect. On the one hand, the studies that clearly find that more features, especially churn-related features, lead to better prediction results, were all conducted on relatively small data sets, i.e., on 26 releases~\citep{Ostrand2005}, three releases~\citep{Moser2008}, and five releases~\citep{DAmbros2012}. On the other hand, \cite{Zhang2017} used 255 and only found a small effect of using aggregations. \cite{Rahman2014} even found no statistically significant difference, when they added features based on PMD or FindBugs warnings to static metrics. Thus, the expectations that larger feature sets that go beyond static source code metrics improve predictions may be inflated. 

\subsection{Prediction Performance}

We note that the overall performance of the model by \cite{CamargoCruz2009} was relatively bad. This is, from our point of view, nearly as troubling as the large amounts of mislabels, because there are two potential interpretations for this bad performance. 1) The benchmark by \cite{Herbold2017} misjudged the performance of the approach by \cite{CamargoCruz2009} based on the prediction performance on five data sets and the approach should have been ranked lower. 2) The benchmark judged the performance correctly, and other approaches from the state of the art perform even worse. Neither interpretation is good. The first interpretation indicates a severe threat to the external validity of existing defect prediction studies, because most of them used the same data as \cite{Herbold2017}. The second interpretation means that release-level defect prediction may not be able to reliably save costs. Regardless which is the case, further research in this direction is important. A potential explanation for the low performance and also possible deviations from the existing defect prediction results would be the relatively small number of defects, in comparison to all existing data set. The rigorous cleaning of false positive defect labels leads to a stronger class-level imbalance in our data, than in any of the existing data sets. Thus, newer results that were not considered in the benchmark by \cite{Herbold2017}, e.g., with respect to sampling for class imbalance treatment~\cite{Tantithamthavorn2018} may help.

\subsection{Open Problems}
\label{sec:open-issues}

While we empirically explore many problems regarding data for defect prediction, there are still open problems left, as well as new problems we discovered due to our results. 

We have only used manual validation for the bug fixing commits, but not for the file actions in those commits or for the inducing changes that were detected. While we used smaller samples to get insights into the quality, this only helps us estimate the remaining noise in the data, as our tooling cannot identify which changes to source code actually contribute to a bug fix automatically. Thus, the first open problem is to extend this data with validated file actions for bug fixes, inducing changes, and release assignments. For the data in our empirical study we would need to manually validate 46.422 file actions for 10.515 bug fixing commits, as well as for the release assignment of 6.530 bugs. These are 34.5 times more file actions than in the study by \cite{Mills2018} with the additional effort for validation of the inducing changes. Thus, this kind of problem cannot be solved by single research groups, but must be tackled by the complete community. We already registered a study with the goal to start to address this problem with the help of the community~\citep{Herbold2020}. In a first step, we want to validate which lines in bug fixes actually contribute to the bug fix and which changes are unrelated improvements. 

Our IND approach for the assignment of bugs to releases also has limitations. For example, lines that are added and not modified do not have an inducing change. In general, as \cite{Perez2020} point out, there are intrinsic bugs, for which they could not locate the inducing change, even with the help of experts. Future research should consider the impact of this limitation on automated approaches, i.e., how often this is the case. Moreover, we should explore if we can improve our heuristics for indentifying inducing changes or if they may even be impossible to solve as an instance of the halting problem~\citep{Turing1937}.

Moreover, our results raise several interesting and concerning questions for further research. We can only speculate how our results regarding the defect labels affect the state of the art. We may find that the same prediction models as before are the best, simply because they are good models, independent of the data. The results may also change, because with the different data, other algorithms may perform better. We are especially looking forward to how our data affects findings that trivial baselines may outperform machine learning, e.g., by \cite{Zhou2018}. Recent work already considered similar problems with other variants of SZZ, but without accounting for the biggest source of noise, i.e., the mislabeled issues~\citep{Fan2019}. The results indicate that these changes will have an effect. 

Through a small experiment, we have already (inadvertently) shown that some results regarding the importance of features may need to be revisited. While we found improvements, they were not significant. However, because we only performed a relatively simple study, this only means that the impact is not as obvious as we expected. Future work may uncover subsets of our features which lead to bigger improvements or demonstrate that we only found negligible differences due to our use of the approach by \cite{CamargoCruz2009}. Even if future research finds that there really is not much of an improvement if other features than static metrics are used, the question of which features are best is still interesting. For example, just-in-time defect prediction relies mostly on features that are independent of the programming language. Whether the same would be possible for release-level defect prediction without loosing predictive performance is unknown. Such results could help to broaden the scope of future research, because current research only considers a relatively small set of programming languages. Vice versa, just-in-time research avoids using static analysis tools and hence, there is a lack of research on the use of features like the complexity of code changes~\citep{Hassan2009}. Future research could explore if the language independent features are really sufficient. 

Finally, there is the impact on mining defect prediction data for industrial application in tools. We rely heavily on time-consuming manual validation to increase the data quality. While we believe that this is essential for researchers that need to accurately and reliably assess proposed approaches, this is of less importance in the industry. Here, the key question is if models trained on noisy data, e.g., collected with SZZ-RA perform as well as models trained on manually inspected data. Our results in Section~\ref{sec:predictions} indicate that this difference may be small. Thus, such data could still be used to train domain specific models. Regardless, our results also indicate that a small sample of data should always be cleaned manually to be used as test data, because the assessment of the model performance may be unreliable. Future research should investigate if our initial results hold and automated heuristics can be safely used for the training of defect prediction models without a negative effect on the prediction performance. 

\section{Threats to Validity}
\label{sec:threats}

Due to the scope of our empirical study, there are many threats to the validity of our findings. We discuss the construct validity, internal validity,  external validity, and reliability as separately, as suggested by \cite{Runeson2008}.

\subsection{Construct Validity}

There are several threats to the construct validity of our experiments. We wrote a large amount of software for these experiments, which may contain defects. However, all software was tested, including the development of automated unit tests for complex and critical components like the identification of inducing changes. Furthermore, we checked the results manually. Especially the large amounts of manual analysis we conducted revealed many corner cases, which we could then handle correctly, mitigating the threats due to bugs in our software.  Additionally, we may have selected unsuitable baselines for the comparison of results. To mitigate this threat, we created ground truth data as baseline where possible. In case this was not possible, we evaluated a sample from our baseline manually to establish whether our proxy for the baseline was suitable. Moreover, we cross-checked all our results with findings from related studies to evaluate the plausibility of our results. Finally, we may have used inadequate metrics for the measurement of differences. We mitigate this threat by only reporting deviations from the ground truth. To further mitigate this threat, we looked at the raw data and validated that the deviations are accurate reflections of the raw results. 

\subsection{Internal Validity}

The results of the analysis of the defect labeling directly follow from the properties of the defect labeling algorithms, e.g., missing links to issues are the only source for false negative bug fixing commits with SZZ in our data. The results of our sampling in Section~\ref{sec:groundtruth-inducing} and the prior work by \cite{Perez2020} both indicate that finding inducing changes based on blaming prior changes may be unreliable. However, the analysis of the differences between the improved versions of this blaming, e.g., by ignoring refactorings, is still valid, because all deviations from simpler approaches are improvements, as we discuss in Section~\ref{sec:inducing-changes}. Similarly, while the assignment of bugs to releases may be negatively affected by this, the results from Section~\ref{sec:groundtruth-inducing} provide a strong indication that these results contain the least amount of noise, in comparison to the other strategies. Thus, while concrete values with respect to the wrong assignment of bugs to releases may change with additional manual validation, the overall conclusions would likely not be affected. 

The conclusion that the difference with a larger set of features is negligible may be wrong or misleading. Other factors, especially properties that we cannot easily capture with performance metrics may yield different results, e.g., the acceptance of prediction models by developers could be higher because the recall is improved. Moreover, we only consider a pure classification scenario and no ranking of files by their likelihood of defect or a regression scenario for the prediction of the number of defects in a file. The additional features may lead to bigger differences in performance under these considerations. 

\subsection{External Validity}

The main threat regarding the external validity of the results is due to our focus on Java projects that are developed under the umbrella of the Apache Software Foundation. Thus, it is unclear if and how our findings generalize to projects using other programming languages or software development outside of the Apache Software Foundation. We note that our analysis regarding the defect labeling problems is mostly independent of the programming language, with the exception of the identification of whitespace and comment-only changes. For example, whitespace changes may actually be changes to the logic of a program written with Python. Moreover, the Apache Software Foundation attracts a large amount of developers both from the industry as well as from the open source community. This increases the likelihood that aspects like the labeling of issues as bug or the use of the affected versions field are similar in other contexts. 

Furthermore, it is unclear if our conclusions regarding performance differences are only relevant for release-level defect prediction or if they generalize to just-in-time defect prediction. However, a follow-up study indicates that there at least large performance differences between keyword-based approaches and validated data for just-in-time defect prediction~\citep{Trautsch2020a}.

\subsection{Reliablity}

To avoid bias in the manual validation of data, we involved multiple people. The issues were validated by two authors independently, conflicts were solved by three authors. While the initial validation of the issue links was conducted only one author, two authors performed the manual analysis of a sample of 1000 issues for mislabels. Additionally, these results were cross-checked by a third person that is not an author of this manuscript. Thus, we minimized the impact of individuals on the results to mitigate this the threat to the reliability of the research. 

\section{Conclusion}
\label{sec:conclusion}

Within this article we performed a critical assessment of the state of practice of the collection of defect prediction data. We summarized existing data sets and found that the SZZ algorithm is the standard approach for defect labeling and that most data sets only offer a limited set of features. This is in contrast to the state of the art that found problems with defect labeling using SZZ, as well as diverse features that should be valuable for defect prediction. To assess the impact of this difference, we performed an empirical study with the focus on the problems of defect labeling and found that SZZ identifies one incorrect bug fixing commit for each correct bug fixing commit, while still missing about one fifth of the bug fixing commits. The main reason for the mislabeled commits are mislabeled issues, a problem initially found by \cite{Herzig2013}. For release-level defect prediction data, this problem is even worse, because most data sets use a six month timeframe to assign defects to releases. The combination of these problems mean that for every correctly labeled defective file, there is one incorrectly labeled defective file and two missed defective files. Thus, there is a large amount of noise in the defect prediction data that is currently used and we can only speculate how this affects the state of the art. Future work on defect labeling should carefully manually validate data and not rely on time windows for the identification of defects to avoid the generation of noisy data.

Regarding the features, we found that the difference of the prediction performance measured with more features is not statistically significant, even though more features yielded the best overall results. This is in contrast to prior findings that highlighted the importance of, e.g., churn features. However, since our analysis of feature importances and the impact of larger feature sets was only rudimentary, additional research is required to establish what a suitable set of features for defect prediction looks like. 

Another contribution of this article is a new defect prediction data set, both for just-in-time, as well as release-level defect prediction. Our data set is larger than any currently used data set, i.e., contains more releases and projects, as well as more features. We hope that the data we produced as part of our work will help the research community to resolve the problems we found. On the one hand, we are looking forward to studies of defect prediction models using our data, both replications of existing work with the de-noised data, as well the the assessment of new approaches and techniques. On the other hand, our data may also be used to improve automated defect labeling, e.g., by trying to automatically correct bug issue labels in issue trackers. Moreover, we hope that our data will be used as the foundation for the manual validation of file actions, to provide a ground truth assessment of the assignment of defects to files and releases. 

\section*{Acknowledgements}

This work is partially funded by DFG Grant 402774445. We also want to thank the GWDG for the support in using their high performance computing infrastructure, that enabled the collection of the large amounts of software metric data. 

\bibliography{literature}

\appendix

\section{Summary of Acronyms}

Throughout the article, we use a many acronyms to refer to different variants of data creation. We summarize these acronyms here. 

\begin{tabular}{lp{7cm}}
SZZ & The original SZZ algorithm by \cite{Sliwerski2005} that identifies by fixing commits through numeric links to issues and uses blames to identify inducing changes.\\
SZZ-RA & An improved SZZ algorithm by \cite{Neto2018, Neto2019} that ignores whitespaces and refactorings when identifying inducing changes. We further extended SZZ-RA to ignore test and documentation changes.\\
JL & Identification of bug fixing commits through the Jira identifier of issues. \\
JLM & Extends JL with manual validation of the links from commits to issues. \\
JLMIV &  Extends JLM with the manual validation of the type of issues to determine is they are really bugs.\\
JLMIV+ & Extends JLMIV with the filtering of documentation, test, and whitespace changes.\\
JLMIV+R & Extends JLMIV+ with the filtering of refactorings. Can be seen as SZZ-RA with Jira links and manual validation of issue links and issue types. \\
6M-SZZ & SZZ with a six months time window to assign bugs to releases. \\
6M-SZZ-RA & SZZ-RA with a six-month time window to assign bugs to releases. \\
6M-JLMIV &  JLMIV with a six month time window to assign bugs to releases.\\
AV-JLMIV & JLMIV with affected versions of linked issues to assign bugs to releases. \\
IND-JLMIV+R & JLMIV+R with inducing changes to assign buts to releases. \\
IND-JLMIV+R-ALL & Release-level defect prediction data with IND-JLMIV+R for bug labeling and all our features.\\
IND-JLMIV+R-SM &  Release-level defect prediction data with IND-JLMIV+R for bug labeling and only static product metrics as features.\\
6M-SZZ-ALL & Release-level defect prediction data with SZZ for bug labeling and all our features. IND-JLMIV+R labels are used for testing. \\
6M-SZZ-SM & Release-level defect prediction data with SZZ for bug labeling and only static product metrics as features. IND-JLMIV+R labels are used for testing. \\
6M-ALL-SZZ-SZZ & Release-level defect prediction data with SZZ for bug labeling and all our features. SZZ labels are used for testing. \\
6M-SM-SZZ-SZZ & Release-level defect prediction data with SZZ for bug labeling and only static product metrics as features. SZZ labels are used for testing. \\
\end{tabular}

\section{Details of the Data Collection}
\label{sec:collection-details}

We used the tools vcsSHARK\footnote{All *SHARK tools and mynbou are available at https://github.com/smartshark}, mecoSHARK, coastSHARK, changeSHARK, and refSHARK to collect data from the version control system. The vcsSHARK collects meta data about commits, e.g., the messages, the committer, as well as the actual changes, i.e., the file actions and hunks. The mecoSHARK is a wrapper around  SourceMeter\footnote{https://www.sourcemeter.com/}, a tool that calculates static software metrics, clone metrics, as well as the warnings by the static analysis tool PMD. The coastSHARK collects AST node counts and the import statements of Java classes, i.e., low level data about the use of language constructs and the dependencies of classes. The changeSHARK is a wrapper around the ChangeDistiller~\citep{Fluri2007} that determines the types of changes performed in commits using the classification from \cite{Zhao2017}. The refSHARK is a wrapper around the RefDiff tool for refactoring detection~\citep{Silva2017}. These tools are executed for every commit in the repositories to collect data about the source code evolution. Additionally, we use the tool memeSHARK to remove redundancies from the collected data, e.g., because the data did not change between commits, for a more efficient storage. We use the tool issueSHARK to collect data from the issue tracking system of the projects, e.g., the identifiers, comments, status, and other meta data about the issues. 

The tools linkSHARK, labelSHARK, and inducingSHARK implement the approaches for issue linking, labeling of bug fixing commits, and the inference of inducing commits\footnote{We use git blame for the identification of inducing commits.} that we evaluate in this empirical study. The manual validation was supported by the visualSHARK, a web application that presented the data that requires manual validation to the experts and stores the results of the validation in the MongoDB. The information that the web interface provides is similar to the LINKSTER tool~\citep{Bird2010}.

Data for just-in-time defect prediction is available through the MongoDB. This data includes all metrics mentioned above, and also the data required for code ownership analysis~\citep{Bird2011}. We use the tool mynbou to create CSV files with release-level data with files as level of abstraction. For each file, the data contains the software metrics and PMD warnings collected by the mecoSHARK and the coastSHARK, the number of the different kinds of changes and refactorings collected by the changeSHARK and the refSHARK from the last six months, and churn metrics proposed by \cite{Moser2008}, \cite{Hassan2009}, and \cite{DAmbros2012}. Additionally, mynbou provides all thirteen aggregations that were proposed by \cite{Zhang2017} for the software metrics the mecoSHARK collected that are not on the file level, i.e., class, method, interface, enum, attribute and annotation metrics. The data set contains a total of 4198 features. We decided not to add features with code smells~\citep{Palomba2019} to the data set, because these can be calculated indirectly from the available source code metrics and definition of smells like god class may change over time. Additionally, features based on mutation testing are not available, because retrospective execution of tests is, unfortunately, often not possible~\citep{Tufano2017}. We also have not added features regarding test smells~\citep{Spadini2018} and review activity~\citep{Thongtanunam2016}, because current results only show correlation with defects, but have not yet shown that these features may actually improve defect prediction models.

\end{document}